\documentclass{emulateapj}
\usepackage{lineno}
\usepackage{natbib}
\usepackage{xcolor}
\usepackage{graphicx}

\newcommand{\project}[1]{\textit{#1}}
\newcommand{\TESS}{\project{TESS}}

\newcommand{\Gaia}{\project{Gaia}}
\newcommand{\dd}{\mathrm{d}}
\newcommand{\rvchisqpval}{\texttt{rv\_chisqu\_pvalue}}
\newcommand{\rvrenormalisedgof}{\texttt{rv\_renormalised\_gof}}
\newcommand{\paired}{\texttt{paired}}
\newcommand{\semiamplitudeprimary}{\texttt{semi\_amplitude\_primary}}

\usepackage{mathrsfs}
\usepackage{comment}
\usepackage{gensymb} 

\usepackage[T1]{fontenc}
\usepackage{ae,aecompl}
\usepackage{breakurl}

\usepackage{amsmath}
\usepackage{amsfonts}
\usepackage{amssymb}
\usepackage{comment}

\usepackage{spverbatim}
\usepackage[%
        ]{hyperref}
\usepackage[nameinlink]{cleveref}

\begin{document}


\title{\paired: A Statistical Framework for Detecting Stellar Binarity with \Gaia\ RVs. \\
I. Sensitivity to Unresolved Binaries}



\author{Quadry Chance\altaffilmark{1,2,$\dagger$}}
\author{Daniel Foreman-Mackey\altaffilmark{2}}
\author{Sarah Ballard\altaffilmark{1}}
\author{Andrew R.\ Casey\altaffilmark{3,4}}
\author{Trevor J.\ David\altaffilmark{2}}
\author{Adrian M.\ Price-Whelan\altaffilmark{2}}
\affil{}

\altaffiltext{$\dagger$}{qchance@ufl.edu}
\altaffiltext{1}{Department of Astronomy, University of Florida, Gainesville, FL, USA}
\altaffiltext{2}{Center for Computational Astrophysics, Flatiron Institute, New York, NY, USA}
\altaffiltext{3}{School of Physics \& Astronomy, Monash University, Victoria, Australia}
\altaffiltext{4}{Center of Excellence for Astrophysics in Three Dimensions (ASTRO-3D), Australia}

\begin{abstract}

Data Release 3 (DR3) from the \Gaia\ Mission includes radial velocity measurements of over 33 million targets. Among many scientific applications, the overlap of this stellar sample with targeted exoplanet transit survey stars presents an opportunity to understand planet occurrence in the context of stellar multiplicity on a large scale. Yet, any interpretation of occurrence relies upon an understanding of survey sensitivity. While the sensitivity to planets in transit surveys is well understood, a characterization of the sensitivity of Gaia to unresolved binaries is also critical. We describe here a statistical framework called \paired, which we developed to enable the forward modeling of \Gaia\ radial velocity observables for large samples of stars. The \paired\ machinery links the reported radial velocity noise for a given star from \Gaia\ to the probability of a spatially unresolved stellar companion.   We demonstrate how this enables the user, given an observed distribution of individual binary ``probabilities" for a set of stars, to understand this distribution within the sensitivity limits of \Gaia. For the subset of stars with the highest probability of excess radial velocity noise, we describe the ability of \paired\ to constrain the semi-amplitude of the stellar binary. Where possible, we benchmark our inferred radial velocity semi-amplitudes against those from ground-based radial velocity surveys, and the subset published by \Gaia\ DR3 itself. We aim for \paired\ to be a community tool for the exploration of the effects of binarity on planets at a population level, but also for any user interested in stellar populations. 
\end{abstract}


\section{Introduction} \label{sec:intro}

The connection between radial velocity uncertainty and stellar multiplicity is well-known \citep{2005maxted,Clark_2011,Maoz_2012}. As RV surveys have evolved to observe hundreds of thousands to millions of stars within the last two decades, the method of using noisy, sparse radial velocity data to statistically characterize stellar multiplicity has evolved alongside them \citep{price2017joker, Price_Whelan_2020}. ESA's \Gaia\ mission, with both astrometric and radial velocity observations, presents a novel opportunity to study binarity on a large scale \citep{gaia_mission_2016,gaia_release_2023}. For example, \cite{el-badry_million_2021} identified 1.3 million spatially-resolved binaries from the \Gaia\ catalog at the $>$90\% confidence level. \Gaia\ DR3 has a catalog of more than 400k binaries with Keplerian orbit solutions and another nearly 400k from spectroscopic trends and astrometric accelerations \citep{gaia_mult}, along with a catalog of more than 2 million eclipsing binary candidates \citep{gaia_eb_2022}.

This type of analysis is timely for exoplanet science, as the increasing number of planets places increasing pressure on traditional time- and resource-intensive campaigns with adaptive optics \citep{Ziegler_2018,ziegler_soar_2021} or radial velocity follow-up \citep{teske2021magellan, chontos2022tess}. For the study of exoplanets, \Gaia\ presents a twofold opportunity to extract binarity information. Firstly, \Gaia\ observes \textit{non}-planet-hosts in the same manner as planet host stars: such comparisons are ultimately crucial for discerning the effects of stellar multiplicity upon planet occurrence at a demographic level. However, the former sample is typically two orders of magnitude larger than latter.  Observations to identify binary companions from the ground often focus upon known planet host stars out of necessity, to eliminate false positives or identify the effects of stellar contamination on inferred planet radius \citep{2015Ciardi}. A survey sample with a common set of observations for planet-hosts and non-planet-hosts more readily enables a comparison between the two populations. Secondly, with respect to planet hosts, it observes the sample in a uniform way. Searches for stellar companions by ground-based  means (whether radial velocity or imaging) are shaped individually by survey completeness, which is itself informed by spectral resolution, stellar magnitude, and observing cadence. This naturally renders some portions of companion parameter space (mass ratio and orbital separation) better studied for some stars in the host star sample stars than others. 

The application of the \Gaia\ DR3 data release to robustly investigate the effects of stellar binarity upon planet formation, however, also requires an estimation of the underlying sensitivity to binaries. This is a non-trivial consideration, given that individual epochal radial velocities are presently available only for a small fraction of the 33.8 million sources observed by the Gaia Radial Velocity Spectrograph \citep{katz_2022}. Among these, only about 1800 RR Lyrae and Cepheids variable stars have individual published RVs \citep{Ripepi23}. An  additional 800,000 sources are flagged as binaries, either due to an observed spectroscopic trend, an astrometric acceleration, or a full Keplerian orbital solution \citep{gaia_mult}. For the remaining 30 million sources (those with Gaia Radial Velocity Spectrograph magnitude $G_{\textrm{RVS}}\le$14 and effective temperature $3100 \le T_{\textrm{eff}}\le14500$), the user presently has access to 3 secondary data products: (1) the median radial velocity \verb|radial_velocity|, (2) the number of radial velocity observations \verb|rv_nb_transits| (where in general, one ``transit'' in the RVS field-of-view includes 3 RVS CCD observations, per \citealt{dr3_documentation_ch20}), and (3) the estimated error on the median, \verb|radial_velocity_error|. This third quantity, denoted by $\epsilon$ and denoted the ``formal uncertainty" on the combined radial velocity measurement, is a function of the standard deviation of the epochal time series \citep{katz_2022}. 

It is this ``error" term that potentially encodes information about binarity for the vast majority of the 30 million stars with radial velocity measurements in DR3. With the open-source Python package \paired, we present a formal framework for leveraging it on a large scale. The machinery of \paired\ is  based upon a probabilistic model to extract binary likelihood for every star with radial velocity observations from \Gaia. The idea of employing excess noise in \Gaia\ data as a diagnostic for binarity (whether in astrometry or radial velocity observations) is not a new one. \cite{Evans18} first identified the use of excess astrometric ``error" in identifying binary candidates in DR2: in that case, the authors employed the keyword \verb|GOF_AL|, corresponding to the ``astrometric goodness of fit in the along-scan direction" \citep{Lindegren18}. They also found the astrometric excess noise term \verb|D|, the significance of the additional uncertainty assigned to measurements of a source to encompass modelling errors, to be correlated with binarity. \cite{Laos20} found that those keyword values correlated strongly with the presence of binary companions within 1.5 mas in adaptive optics imaging. \cite{Belokurov20} used instead the ``renormalized unit weight error", analogous to the reduced $\chi^{2}$ of the astrometric fit (Renormalised Unit Weight Error(RUWE), \citealt{Lindegren18}) as a proxy for stellar multiplicity. They identified that two regions of the Hertzsprung-Russell diagram contained strong RUWE excess: the multiple-star Main Sequence that lies about the single-star Main Sequence, as well as the white dwarf-M dwarf binary sequence. Other investigations have employed excessive \Gaia\ noise, whether in astrometric or radial velocity keywords, to identify or cull likely binaries with the same reasoning \citep{Penoyre_2022a, Penoyre_2022b, Wolniewicz21,Lu22, Chontos22, Anderson21, Berger20a, Berger20b, Bryson20, Shabram20, Hsu19}.

The \paired\ pipeline and catalog is intended to be an ancillary tool to the \Gaia\ releases of radial velocity data. We crafted \paired\ to be capable of forward-modeling Gaia radial velocity parameters that are (a) useful for the identification of binarity on a large scale and (b) currently available to the public (to enable comparison between prediction and observation). The volume of information furnished by the Gaia data releases is enormous, and we designed \paired\ to complement it by meeting three presently outstanding scientific needs. 

\begin{enumerate}
    \item Firstly, \paired\ functions as an open-source translator, allowing the user to map physical properties of binary star systems (by which a user defines mass ratio, orbital separation, inclination, magnitude, etc.) to their corresponding predicted ``noise" properties in the Gaia DR3 catalog of sources observed by the RVS \citep{katz_2022}. In the case of unresolved binaries, without the individual epochal measurements, binarity is detectable only via seemingly excess radial velocity noise. Our aim is for \paired\ to identify when this noise is detectably ``excessive", and to what extent, in a way that is readily linked to the underlying stellar population. For about 6 million targets, DR3 publishes analogous quantities, corresponding to a likelihood of excessive noise \rvchisqpval\ and a goodness-of-fit parameter between the observed RVs and a single-star model \rvrenormalisedgof. For these 6 million targets, we can compare the result of our open-source machinery to published Gaia analogs. 
    
    \item Employing \paired, we explore \Gaia's sensitivity to physical binary properties with an injection-and-recovery analysis, and thus estimate the ``completeness" to unresolved binaries. The estimation of \Gaia's radial velocity sensitivity to binaries as a function of the mass ratio, orbital separation, and magnitude of the primary star is, unto itself, a useful tool for occurrence rate studies; \cite{Penoyre_2022a} performed a similar sensitivity analysis for unresolved binaries detectable by excess astrometric noise. 
    
    \item For the remaining 23 million stars with published \verb|radial_velocity_error| but presently no \verb|rv_chisq_pvalue| (that is, those with $G_{\textrm{RVS}}>$14), we have employed \paired\ to extract the likelihood of ``excessive" radial velocity noise. For these sources, there is presently no published analog in DR3. 
    
\end{enumerate}
The pipeline for creating a custom catalog with \verb|paired| can be found \href{https://github.com/quadrychance/paired}{at this URL}.
The catalog used in this work is hosted on the open source data repository Zenodo \href{https://doi.org/10.5281/zenodo.7749554}{at this URL}. It contains a table with the following columns:
\begin{enumerate}
\item  Information intended to assist in crossmatching \begin{spverbatim}'source_id','ra','dec', 'parallax'\end{spverbatim}

\item B-R color and apparent G magnitude \begin{spverbatim}'bp_rp', 'phot_g_mean_mag'\end{spverbatim}

\item Gaia keywords associated with binarity \begin{spverbatim}'rv_renormalised_gof', 'rv_chisq_pvalue', 'rv_amplitude_robust'\end{spverbatim}

\item The estimated per-transit measurement uncertainty and associated error \begin{spverbatim}'rv_ln_uncert', 'rv_ln_uncert_err'\end{spverbatim}

\item The p-value for each target with Gaia DR3 RV in the sample under the null hypothesis of a single-star time series \verb|'rv_pval'|

\item The 5th, 16th, 50th, 84th, and 95th percentile constraints on the RV semiamplitude derived from the RV noise assuming Keplerian motion \begin{spverbatim}'rv_semiamp_p5', 'rv_semiamp_p16', 'rv_semiamp_p50', 'rv_semiamp_p84', 'rv_semiamp_p95'\end{spverbatim} Note: this column is only calculated for sources with p-value < 0.01
\end{enumerate}

  This manuscript is organized in the following manner. In Section \ref{sec:methods} we describe the methodology of \paired\ for extracting the likelihood of excess radial noise from the publicly available radial velocity \Gaia\ data. We assess our sensitivity to stellar companions as a function of mass ratio, orbital separation, and apparent magnitude with a series of injection-and-recovery tests. In Section \ref{sec:results}, we run \paired\ on a sample of $\sim$30 million \Gaia\ sources observed with the RVS, and describe our findings. We compare the $p$-values generated by \paired\ to the analogous \rvchisqpval\ and \rvrenormalisedgof\ provided for a subset of the \Gaia\ DR3 sample. We also investigate how our sample of ``likely binaries" identified by \paired\ overlaps with samples of binaries identified by other (non-\Gaia) means. This includes the presentation of eclipses in photometry and the scatter as measured from from ground-based radial velocity surveys. In Section \ref{sec:conclusion}, we summarize our findings and suggest a number of future potential use cases for \paired.

\section{Methods} 
\label{sec:methods}

The machinery of \paired\ relies upon a simple question: does a source have RV noise in excess of the amount expected for similar-looking single stars? The RVs of a single star should have Gaussian noise with a standard deviation set by the observational uncertainty and any underlying astrophysical sources of noise. Since the amount of intrinsic stellar RV variation for stars with convectively or magnetically driven RV jitter is generally too low ($<$100 m/s) to be detectable with the Gaia Radial Velocity Spectrograph \citep{Luhn_2020,brems_2020}, the individual measurements for a single target should therefore be normally distributed, with a standard deviation dominated by the measurement uncertainty. If the criteria for identification as a binary is set conservatively at 3-$\sigma$, in theory, few single stars should be identified as binaries.  Of course, for sources with otherwise larger physically-driven RV variability such as young stars, particularly active cool dwarfs, oscillating giants and sub-giants, and rapidly rotating hot stars, ``excess" radial velocity noise is harder to quantify.  We aim for our method to be robust to these variations until their variability amplitude reaches parity with the estimated per-transit measurement uncertainty for that particular source. The details of this operation are expanded upon in Appendix \ref{sec:appendix-chi2}.

In addressing the question of whether a given source exhibits RV noise in excess of the amount expected for similar-looking single stars in \Gaia, we employ the presently publicly available (as of DR3) measurements: (1) the median radial velocity, (2) the uncertainty on this median radial velocity, and (3) the number of \Gaia\ radial velocity observations used to quantify (1) and (2). Figure \ref{fig:RV_transits} depicts the distribution in the number of transits, for the 33.8 million sources observed by the Gaia Radial Velocity Spectrograph \citep{katz_2022}. The average number of individual radial velocity epochs that contribute to the calculation of the reported median RV and its uncertainty is 20. These reported median and uncertainty measurements are derived under the assumption that the measured radial velocities are attributable to a single star. The degree to which a single star model will furnish a particularly ``poor'' fit, rendering the binary identifiable, is dependent upon several factors. These include the time sampling of the radial velocity curve, the properties of each individual star, and the properties of the binary itself. These latter properties include mass ratio, orbital separation, eccentricity, and alignment relative to the line of sight; along with the mass, spectral type, and evolutionary state of the components. In Section \ref{sec:calc_radvel} we detail our procedure for identifying anomalously high radial velocity errors, which is based upon a comparison among similar stars. In Section \ref{sec:injection} we characterize our sensitivity to binaries as a function of mass ratio, orbital separation, and stellar magnitude. We determine the part of parameter space for which binaries are identifiable from excess RV noise in Section \ref{sec:recover_excess}, and demonstrate our ability to then reverse-engineer the semiamplitude of those binaries in \ref{sec:recover_K}. In Section \ref{sec:false-positives}, we consider potential sources of false positives. 




\subsection{Formalism of \paired}
\label{sec:calc_radvel}

Appendix \ref{sec:appendix-chi2} describes our method of binary detection and characterization in more detail, and we summarize the key results here. As discussed above, the \Gaia\ catalog includes an estimate of the ``RV error'',  \verb|radial_velocity_error|, for each target with Gaia DR3 RV. Since this error is estimated by computing the sample variance of the RV time series \citep{katz_2022}, it encodes information about both the RV measurement uncertainty and the excess RV variability introduced by companions. To identify probable binaries in this sample, we build a data-driven model to estimate the ``typical'' per-observation RV measurement uncertainty as a function of color ($B_{P}$-$R_{P}$) and apparent magnitude ($G$ band). We can then use this estimate to identify targets with ``atypical" statistically significant excess RV error.

To assess whether a source exhibits RV noise in excess of the amount expected for similar-looking single stars, we first partition the observed \Gaia\ catalog stars into bins in color-magnitude space. For each of these bins, we tabulate the average radial velocity error. The \Gaia\ DR3 catalog includes RV measurements made using two different methods, one for targets brighter than $G_\mathrm{RVS} = 12$, and a different one for fainter targets \cite{katz_2022}. For the bright targets, the RV error is estimated based on the sample variance of the RV time series. For the targets fainter than $G_\mathrm{RVS} = 12$, the RV error is computed  \textcolor{green}{from the sharpness of the combined cross-correlation function} described in \cite{katz_2022}. The nature of this scatter will vary from bin to bin: some stars will have intrinsically higher amounts of RV noise for reasons unrelated to binary companions. In other portions of color/magnitude space, this sample is more intrinsically contaminated by binaries. Calculating the average radial velocity error for that bin therefore requires an iterative clipping process (see Appendix \ref{sec:appendix-noise} for more details), whereby we exclude the noisiest sources in our calculation.

We then compute a $p$-value for each target with Gaia DR3 RV in the sample under the null hypothesis of a single-star time series. Informally, for a given target, this $p$-value quantifies the probability that such an extreme RV error would have been measured if the system were in fact a single-star system with measurement uncertainty. Somewhat more formally, for a target $k$, this $p$-value can be written as
\begin{equation}
    \label{eq:$p$-value}
    p_n = \mathrm{Pr}(\xi > \xi_k\,|\,\mathrm{target}\,k\,\mathrm{is\,single})
\end{equation}
where $\xi_k$ is the observed RV error for target $k$.
It is important to note that this is not the same as quantifying the probability that a given target is a binary but, as we show below, it is a useful quantity for selecting candidates. 

Where \verb|radial_velocity_error| is estimated based on the sample variance of the RV time series (that is, for $G_\mathrm{RVS}<12$), our $p$-value metric in Equation \ref{eq:$p$-value} can be computed in closed form as discussed in Appendix \ref{sec:appendix-chi2}. When \verb|radial_velocity_error| is computed from the sharpness of the combined cross-correlation function (that is, for $G_\mathrm{RVS}>12$), the $p$-value metric can no longer be computed as simply. We do, however, use the same expression to evaluate Equation \ref{eq:$p$-value} in both cases and find that the results are robust when compared to known spectroscopic binary catalogs (Section \ref{sec:compare_RV}).

In the following sections, we identify probable candidates by selecting targets with $p$-values below $0.001$. This is a relatively conservative definition of ``outliers" at the $>$3-$\sigma{}$ level from the mean in each bin. While we can be reasonably certain that our detections represent significant RV variations, we also understand that such a stringent $p$-value criterion will leave us with a significant false-negative rate. We consider our recovery sensitivity in Section \ref{sec:injection}. 

For likely binary candidates, we also place probabilistic constraints on the RV semi-amplitude $K$ for each target, under the assumption that the RV variations can be described by Keplerian motion. The details for this procedure are described in Appendix \ref{sec:appendix-semiamp}. This $K$ value can be compared to the reported \verb|semi_amplitude_primary| in the DR3 catalog. Though \paired\ calculates the former from the RV variance alone, and the latter is calculated with individual radial velocity epochs, both quantities refer to the predicted semi-amplitude.     


\subsection{Injection and recovery analysis}
\label{sec:injection}

We calibrate the sensitivity of the method described in Section \ref{sec:calc_radvel} with an injection and recovery analysis. Given a ground truth knowledge of the injected population, we can then assess our recovery performance. We consider two types of ``recovery:" first, whether \paired\ correctly identifies a given binary, by virtue of its ``excess" RV noise at the level of $p<0.001$ (results summarized in Section \ref{sec:recover_excess}). Secondly, we consider the accuracy with which we recover the semi-amplitude $K$ of the primary (results summarized in Section \ref{sec:recover_K}).

\begin{figure}[htbp]
\centering
\includegraphics[width=0.49\textwidth]{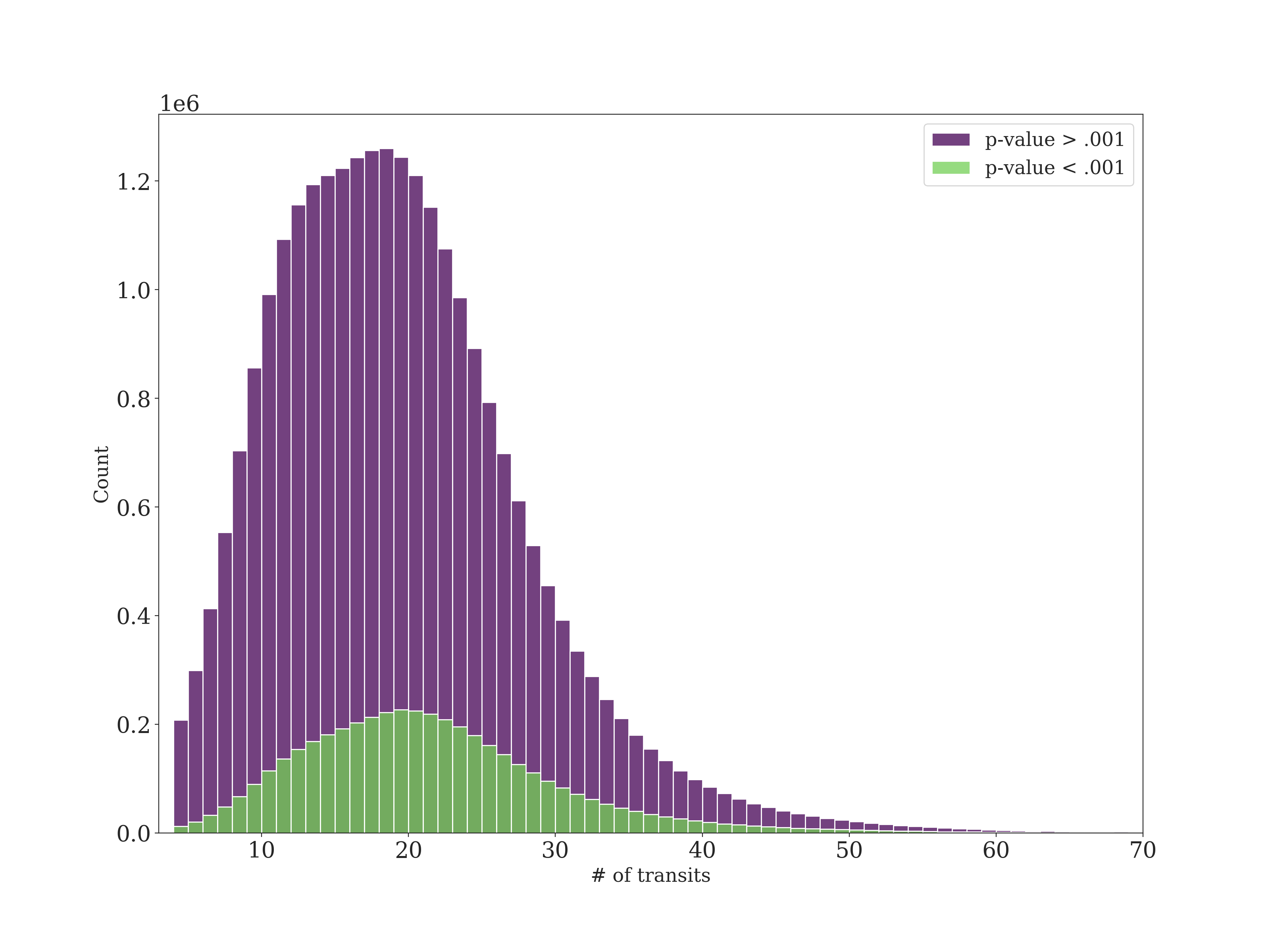}

\caption{A histogram of the number of RV observations (transits) of sources in the \Gaia\ RV sample. \label{fig:RV_transits}}
\end{figure}

We create a synthetic population of binaries for the injection test as follows. First, we draw our underlying stellar binary properties with uniform log spacing in semi-major axis, $a$, and in mass ratio, $q$. Our range in $q$ spans $10^{-4}$ to $1$, which for a Sun-like star brackets a range with a Jupiter-mass exoplanet at one extreme and an equal mass stellar companion at the other. Our semi-major axis range spans $10^{-2} - 10^3$~AU, but our sensitivity falls off substantially for $a>10$ AU. Within this $\{q,a\}$ parameter space, we generate one million binary stars. We then sample and assign orbital inclination relative to the line of sight $\cos{i} \sim \mathcal{U}(0,\,1)$, orbital eccentricity $e \sim \mathcal{U}(0,\,1)$, and longitude of periapse $\omega \sim \mathcal{U}(0,\,2\,\pi)$ from uniform random distributions. Next, we calculate the binary's orbital period and employ Kepler's equations to solve for the positions and velocities of the two stars. The number of synthetic observations $N$ is then drawn from the distribution of RV transits (measurements by the RVS instrument) shown in Figure \ref{fig:RV_transits}. We ``measure" the radial component of the velocity at uniform random times  over the 1038-day observing time span of DR3.  We also calculate the semi-amplitude of the heavier component, assuming it to also be brighter. 

We add realistic per-measurement radial velocity error as follows. We randomly assigned each simulated stellar pair to a color-magnitude bin in the \Gaia\ RV sample, and draw our per-measurement error from the observed underlying distribution for that bin (see Appendix \ref{sec:appendix-noise} for additional details). We then calculate the median radial velocity and radial velocity error of these ``observations" as synthetic versions of \verb|radial_velocity| and \verb|radial_velocity_error|. We can then calculate the resulting $p$-value per Equation \ref{eq:$p$-value}, and for sources that present low $p$-values, estimate the semi-amplitude. 

\subsubsection{Recovery of sources with ``excess" RV noise}
\label{sec:recover_excess}

\begin{figure*}[ht]
\centering
\includegraphics[width=0.8\textwidth]{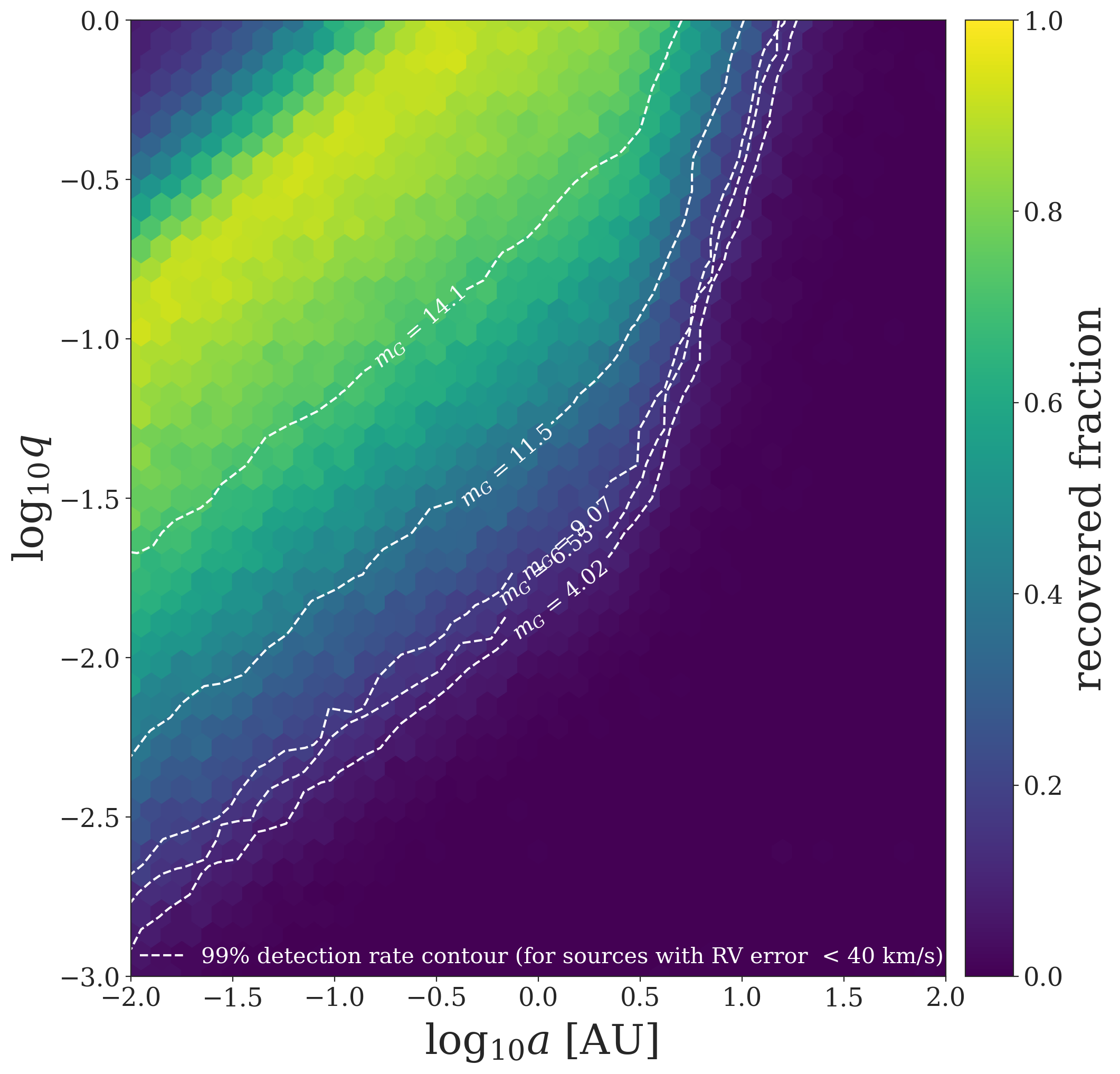}
\caption{Detection efficiency using RV errors as a function of injected separation and mass ratio. 
The color bar is the recovered fraction of binaries over the entire magnitude range of our sample. The contours indicate the threshold where 99\% of binaries to the left of the dotted lines are recovered. This threshold goes to lower mass ratios and larger separations as a function of magnitude. Our recovered fraction decreases in the upper left corner as sources with RV error greater than 40 km/s are culled from the data release.}
\label{fig:detection_efficiency}
\end{figure*}

\begin{figure*}[ht]
    \centering
    \includegraphics[width=0.95\textwidth]{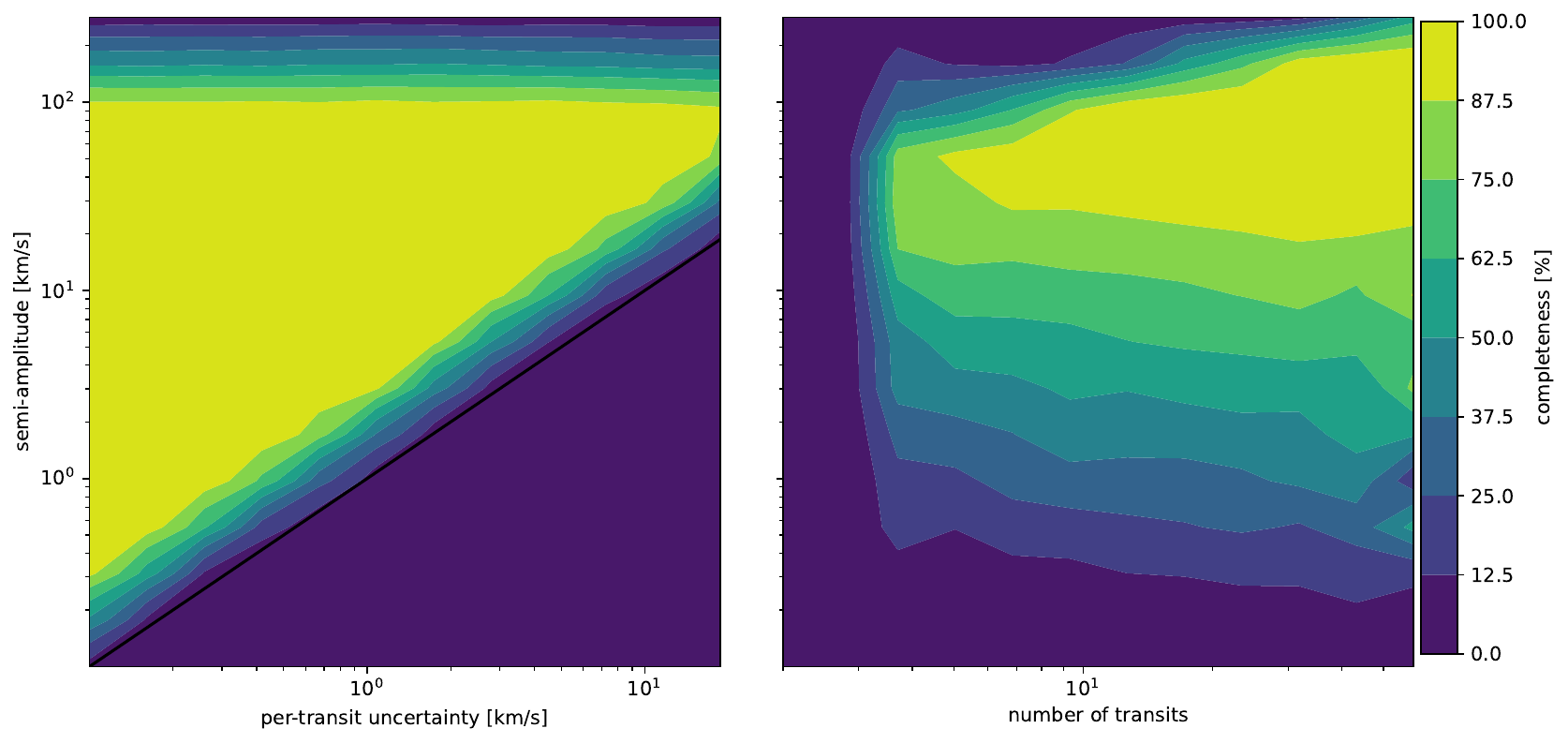}
    \caption{The completeness of our detection method as a function of the observable parameters of simulated binaries. The completeness drops at extreme semi-amplitudes as sources with reported RV errors greater than 40 km/s are culled from the catalog.}
    \label{fig:completeness-observed}
\end{figure*}

We run \paired\ on the simulated sample of binary stars described above, with our results depicted in Figure \ref{fig:detection_efficiency} as a function of semi-major axis and mass ratio. We define ``recovery" as resulting in a \paired\ $p$-value of $<$0.001. The recovery rates with overplotted 99\% recovery rate contours are depicted by white contours corresponding to 5 different \Gaia\ magnitudes. We find that our recovery rate drops off steadily with mass ratio, but more quickly with separation. Predictably, the sensitivity also changes as a function of semi-amplitude, the per transit RV uncertainty  $\sigma$ and the number of RV transits. We show these results in Figure \ref{fig:completeness-observed}. In the left panel, the completeness drops off sharply as $\sigma$ approaches the semi-amplitude of the binary, as our ability to distinguish signal from noise decreases. In the right panel, the sensitivity is mostly a function of the semi-amplitude of the binary with a small increase in completeness as a function of the number of transits after exceeding the threshold of 3 observations.

The results of this injection-and-recovery test indicate that we have some sensitivity to brown dwarfs around bright stars ($m_{G}<7$) out to a few AU ($\log_{10} q = -1.1$ for a Sun-like star). Around those same stars, we are sensitive at the 99\% confidence level to mass ratios of $q$=0.1 at 3
AU. It is outside the scope of this work to compare raw binary occurrence to other works, but we note (per \citealt{Duquennoy91}) that 10 AU for a G dwarf corresponds to an orbital period of $P \sim 10^4$ days, where 90\% of companions to G dwarfs have $q>0.1$. We, therefore, possess modest completeness to stellar companions to bright G dwarfs. 


There is an additional consideration for our injection-and-recovery analysis, related to sensitivity in the portion of $\{a,q\}$ space corresponding to short-period binaries with mass ratios $q$ near to 1. While we would \textit{a priori} expect our sensitivity to peak in this corner of binary parameter space, we note that sources with radial velocity errors $>$40 km/s are culled from the \Gaia\ sample of radial velocity stars and not reported. Equal-mass binaries will also show double-lined spectra. The \textit{Single Transit Analysis} pipeline described in \cite{katz_2022} detects these stars and their RVs are not published. This is reflected in the lower recovery rate at the top-left of \ref{fig:detection_efficiency}. This down-selection necessitates careful consideration of the \paired\ sensitivity map when compared to known samples of binaries.

\subsubsection{Recovery of RV semi-amplitude}
\label{sec:recover_K}

For sources that have $p$-values less than $0.01$, we infer the semi-amplitude through the process described in Appendix \ref{sec:appendix-semiamp}. In Figure \ref{fig:inferred_vs_sim}, we show the resulting inferred semiamplitude, versus the synthetic injected semi-amplitude. 
%
Using the systems created for the injection and recovery test, we can also determine if our method is successful in recovering the correct semi-amplitude and in what conditions our ability to do so is degraded. Our ability to recover the injected semi-amplitude depends mostly on the accuracy of our estimation  of the per-transit measurement uncertainty and the signal-to-noise ratio of the RV signal.
In the first panel of Figure \ref{fig:inferred_vs_sim}, the semi-amplitude is recovered within 1-$\sigma$ when the per-transit measurement uncertainty is exactly known and the signal-to-noise ratio (SNR) is large. The inferred semi-amplitude  for binaries with $e=0$ will be biased slightly high because we are sampling over eccentricity with bounds of 0 to 1 and the heavy tails of the $K$ probability distribution are weighted towards larger $K$. As the SNR approaches unity, our inferred semi-amplitude increases as the noise approaches the level of a real RV signal. This is particularly apparent when, as in the second panel of Figure \ref{fig:inferred_vs_sim} the estimate of the measurement uncertainty is poor. Caution should be taken when the per-transit measurement uncertainty estimate is close to the radial velocity value.

\begin{figure}[h]
\centering
\includegraphics[width=0.48\textwidth]{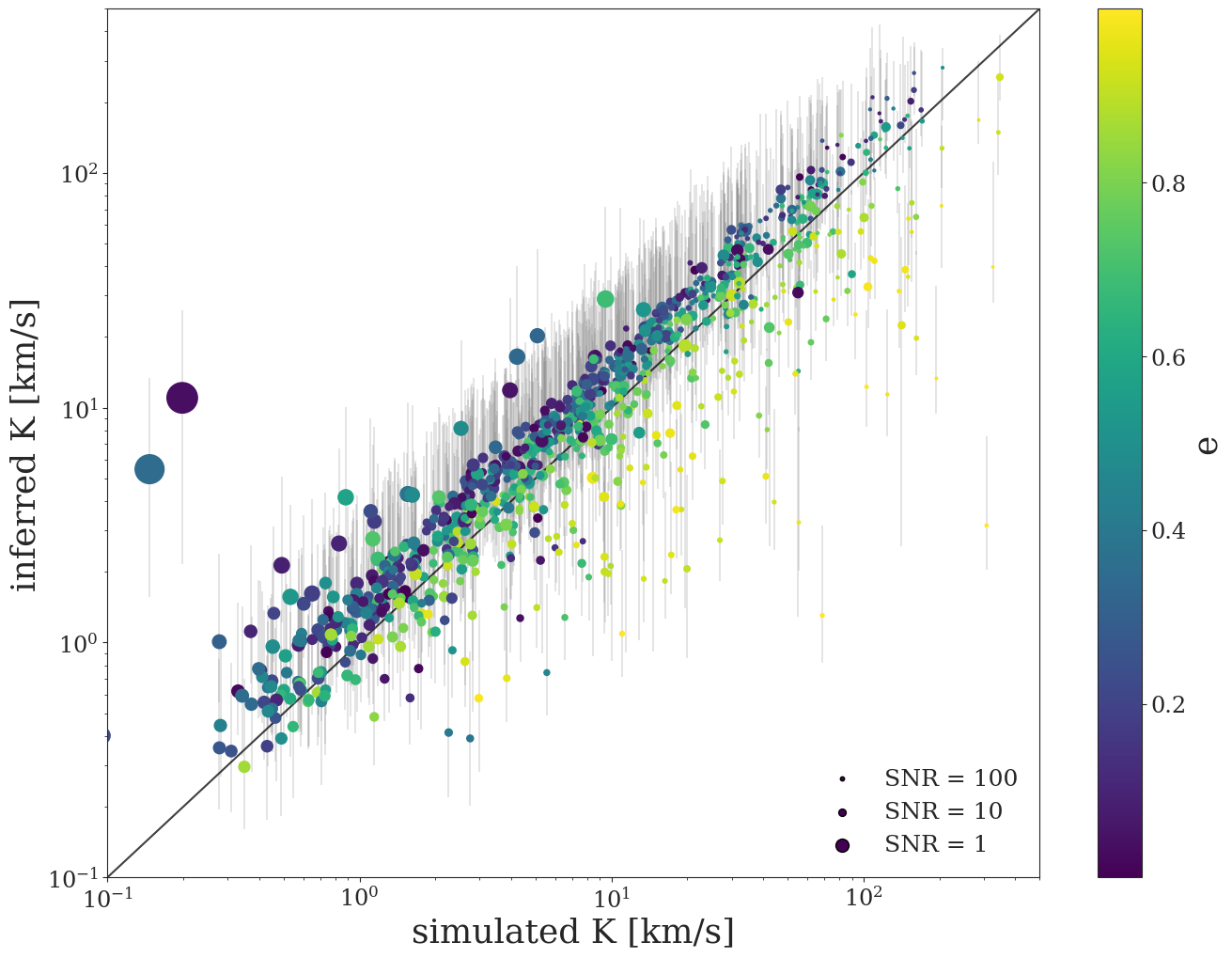}
\includegraphics[width=0.48\textwidth]{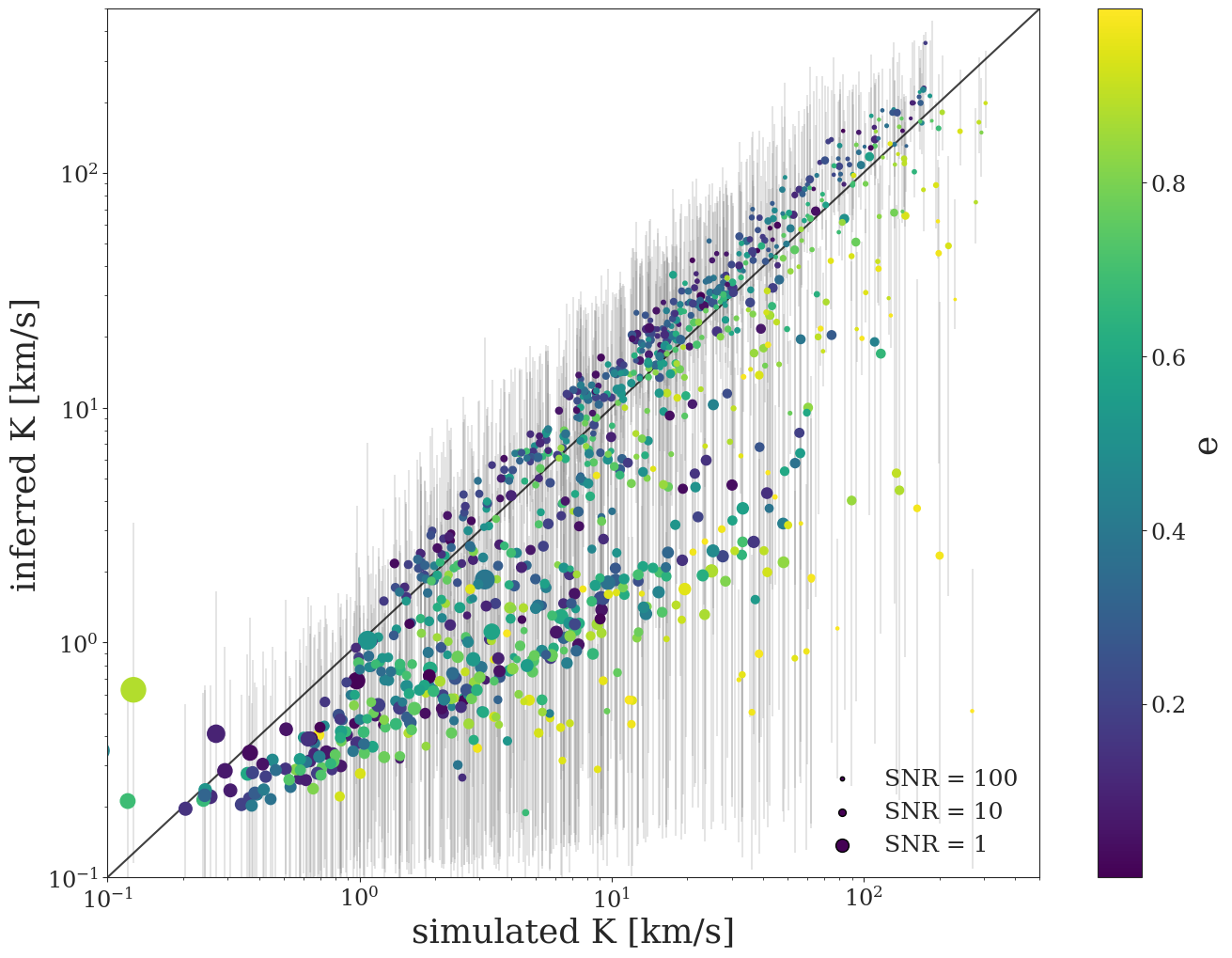}
\caption{{\textit{Top panel: }In an effort to determine if our pipeline robustly recovers the correct $K$, we run it on a population of binary stars described in Section \ref{sec:injection}. We inject the same estimated per-measurement uncertainties for each color and magnitude bin. The size of the circles indicates the ratio of the recovered signal to this uncertainty.  Recovering the correct semi-amplitude mostly depends on how uncertain the RV noise estimate is of stars in the same bin. The recovered semi-amplitude tends to be biased high for two reasons: because we are also sampling over eccentricity with bounds of 0 to 1 and the heavy tails of the $K$ probability distribution are weighted towards larger $K$. The remaining non-random scatter depends on the eccentricity of the binary orbit.} \textit{Bottom panel: } The same as the top panel with a
50\% uncertainty added to the estimated per-transit measurement uncertainty. Sources with a high SNR remain unaffected, but those with low SNR systematically underestimate the semi-amplitude.}
\label{fig:inferred_vs_sim}
\end{figure}

\subsection{Potential sources of false positives}
\label{sec:false-positives}

Previous studies of ``excess" noise as a diagnostic for stellar multiplicity in \Gaia\ have explored the possibility of false positives. In some cases, the reported \Gaia\ RV noise may be underestimated. In other cases, the estimated noise level is correct, but is caused by something other than a binary companion; that is, a physical effect is producing excess RV noise. In this sense, the $p$-value calculation from ``excess" radial velocity noise cannot be interpreted straightforwardly as the false positive rate for our sample across the whole HR diagram in a uniform way. The nature of such false positives, whether due to underestimation of the formal error or \textit{bona fide} excess RV noise from a non-Keplerian source, will likely vary across the Main Sequence. 

We first consider the potential underestimation of \verb|radial_velocity_error|. The DR3 data release from the Radial Velocity Spectrograph \citep{Babusiaux23} investigated the formal \verb|radial_velocity_error|, comparing radial velocities derived by \Gaia\ and APOGEE radial velocity for a set of common sources. If \verb|radial_velocity_error| accurately reflects the measurement uncertainty, then the radial velocity differences between the two surveys should behave in a predictable way: when normalized by \verb|radial_velocity_error|, the $\Delta$RV will be well-represented by a Gaussian distribution with standard deviation of 1. In fact, this is the case for Sunlike stars dimmer than $G_{\textrm{RVS}}$=12. However, \verb|radial_velocity_error| is typically underestimated for sources brighter than $G_{\textrm{RVS}}<12$, with $T_{\textrm{eff}}<$4500 K, or $T_{\textrm{eff}}>$6000 K \citep{Babusiaux23}. However, \paired\ estimates ``typical" RV noise in an empirical manner, by comparing the RV noise for that source to stars of similar apparent magnitude and color. In this sense, if the \Gaia\ RV error is underestimated in a systematic way, other stars at that color and magnitude will be similarly affected, and \paired\ can still identify ``atypical" noise within that sample.   

Contamination by false positives may also occur when the radial velocity error is accurate, but the radial velocity noise is genuinely high for a reason other than orbital reflex motion. This contamination problem was considered for \Gaia\ astrometric noise by \cite{Belokurov20}, who cautioned that a seeming excess in the RUWE can be due to stellar variability rather than binarity \textit{per se}. For this reason, they described, RUWE values for variable stars are difficult to interpret. The RUWE values are additionally elevated for B stars, and the reddest portion of the AGB \citep{Belokurov20}. \cite{Penoyre_2022b} found that in much of the Main Sequence, ``relative flux variability" correlated strongly with astrometric noise. The sub-Main Sequence was the only part of the sample, however, for which the ``spurious" fraction of seemingly high-noise stars was comparable to the binary fraction. These stars, they concluded, exhibit elevated astrometic noise due to a property other than binarity. While we are considering radial velocity noise rather than astrometric noise, similar issues may plague \paired's $p$-values in this part of the Main Sequence. 

Stars with large surface movements from pulsations or convection will also exhibit excess RV noise \citep{Hatzes98,Hekker08, Luhn_2020}. Another known source of false positives specific to radial velocity noise is rapidly rotating stars. The link between projected rotational velocity and RV jitter is a well-known phenomenon \citep{Saar97, Desort07, Boisse11,Hojj20}. In our case, the rotation-broadened spectral lines of stars with $v\sin{i} > 10$ km/s can lead to an added source of noise that will only be partially accounted for when using a group of similar stars that do not all rotate along our line of sight. Rapid rotators will likely have artificially inflated $p$-values, while stars that rotate more slowly (or not along our line of sight) will present smaller $p$-values when compared to the average sample of hot stars. We do not distinguish astrophysical RV noise from a companion, therefore rapid rotators are more likely to be flagged as binaries. 


\section{Results} 
\label{sec:results}

In the previous section, we assess our theoretical sensitivity to binaries with the machinery of \paired. In this Section, we examine our findings after running \paired\ on a sample of $\sim$30 million sources observed by the Gaia Radial Velocity Spectrograph \citep{katz_2022}. Given the 33.8 million sources total, we exclude sources with less than three transits, without which the RV noise cannot be well defined; this excluded sample with $N$=1 or 2 comprises 3.3 million stars. In Section \ref{sec:binary_sequence}, we populate the Hertzsprung-Russell diagram with these $\sim$30 million sources, considering how \paired's $p$-values vary across the Main Sequence and into the evolved stellar populations. In Section \ref{sec:DR3_comparison}, we consider the $\sim$6 million stars (those with $G_{\textrm{RVS}}<$14) for which DR3 has published analogous quantities to those produced by \paired. For these targets, we can directly compare our $p$-value against the \rvchisqpval\ column in the published DR3 catalog. This comparison is described in Section \ref{sec:rvchisqpval}, and while we observe general strong agreement, we also investigate places where our $p$-values deviate. In Section \ref{sec:rvrenormalizedgof}, we consider the \rvrenormalisedgof\ column for these same 6 million stars. We investigate the relationship between \rvrenormalisedgof\ and the corresponding \paired\ $p$-value. In Section \ref{sec:DR3_compare_K}, we focus upon a subset of the $\sim$6 million sources for which DR3 includes a Keplerian orbital solution: this sample includes $\sim300,000$ stars \citep[][Gosset et al.\ in prep]{halbwachs_2022}. We also compare the semi-amplitude published by \Gaia\ for a subset of spectroscopic binaries, against the semi-amplitude we infer from \paired.  

Having considered \paired's results when compared directly to analogous quantities from \Gaia, in Section \ref{sec:Binary_comparison} we go on to validate \paired's performance on a high fidelity sample of known eclipsing and spectroscopic binaries from other catalogs. In Section \ref{sec:paired_EB_samples} we investigate how \paired\ performs on a sample of known eclipsing binaries, and in Section \ref{sec:paired_SB_samples}, we investigate how \paired\ performs on a sample of known spectroscopic binaries-- under what circumstances does \paired\ flag these \textit{bona fide} binary sources as such? For the latter, we can also compare our inferred semi-amplitude $K$ calculated with \paired, to the published value from the survey's epochal radial velocity observations. We describe this comparison in Section \ref{sec:compare_RV}. 



\subsection{Location on the Hertzsprung Russell Diagram}
\label{sec:binary_sequence}

\begin{figure}[ht]
\centering
\includegraphics[width=0.49\textwidth]{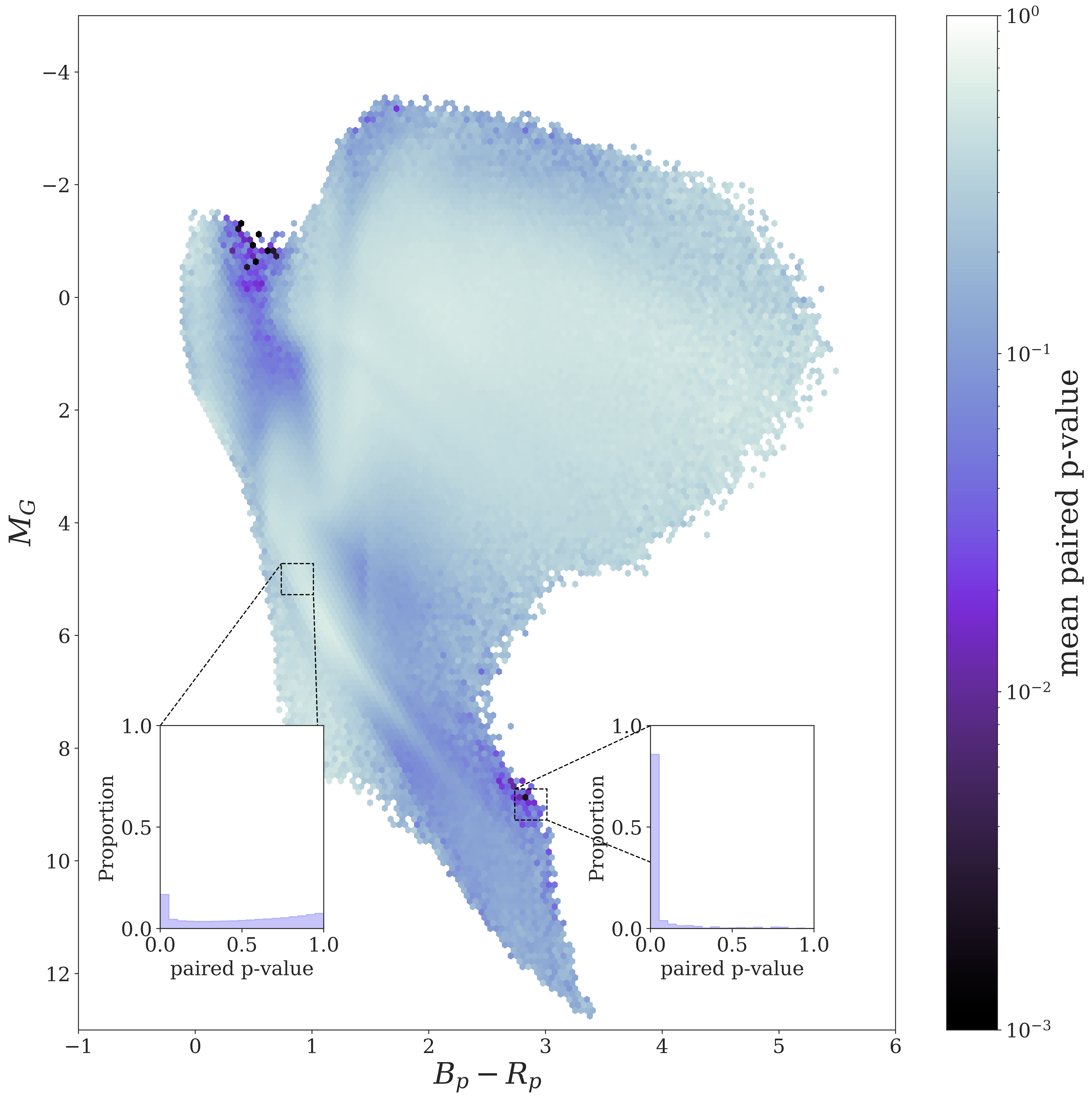}
\includegraphics[width=0.49\textwidth]{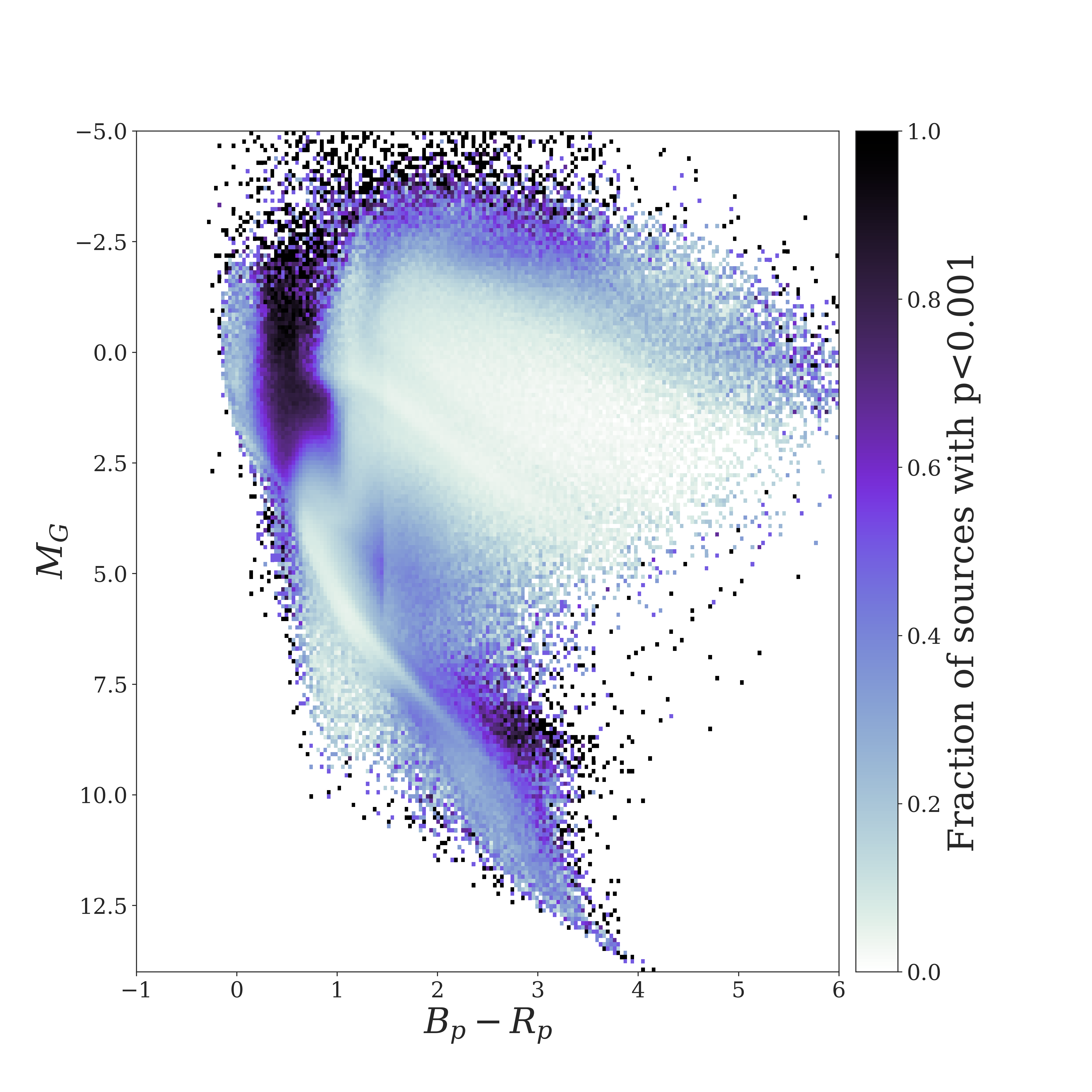}
\vspace{0mm}
\caption{\textit{Top panel: }Hertzsprung-Russell diagram of our sample, comprising $\sim$30 million sources. Colors and absolute magnitudes have not been corrected for extinction. Bins with fewer than 100 sources have been excluded. The color corresponds to the mean $p$-value of the bin, with darker colors indicating a higher fraction of binaries. The inset histograms are the distribution of $p$-values for sources in the marked area. \textit{Bottom panel: }The same sources, with color now assigned based on the fraction of sources with \paired\ $p$-value$<$0.001, which is our cutoff for ``excess" noise. }
\label{fig:HR}
\end{figure}

Figure \ref{fig:HR} maps the resulting $p$-values of 30,545,303 sources onto the Hertzsprung-Russell diagram. We construct a ``likely binary'' catalog by selecting all targets with a $p$-value (described in Section \ref{sec:methods}) below $0.001$. We show the variation in $p$-value in two ways. First, cells in color-magnitude space are assigned shading based on the median $p$-value, with darker bins signifying areas where, on average, the $p$-value is lower. Secondly, we assign shading based on the fraction of sources with ``excess" radial velocity noise. We emphasize that the $p$-value is not the same quantity as the false positive rate of binarity: whether this ``excess" noise is attributable to binarity \textit{per se}, or whether other effects are operative, are questions we consider in this Section. 


We find that $14.4\%$ of the stars observed by \Gaia's RVS instrument lie below this $p$-value threshold, with the percentage varying across the HR diagram. While 10\% of Sunlike stars show excess noise by this criterion (that is,  \paired\ $p<0.001$), the number rises to 20\% for the Young Main Sequence. 
The fraction increases for brighter
sources ($M_{G}<0.0)$, which we expect to be dominated by higher multiplicity systems \citep{Duchene13, Offner23}. We note general agreement on the fraction of sources with excess noise, as compared to \cite{Penoyre_2022a}. In that work, the authors investigated \Gaia\ sources of excess \textit{astrometric} noise (defined in that case as Local Unit Weight Error, or LUWE, $>2$). Within the photometric binary Main Sequence, the fraction of sources showing excess RV noise approaches 100\%. For particularly dim and blue sources on the opposite side of the Main Sequence, the fraction of noisy RV sources also increases. Per \cite{Penoyre_2022a}, these could potentially be Main Sequence/white dwarf binaries. 

The \paired\ $p$-values do exhibit an interesting departure in the lower mass and dimmer part of the Main Sequence (where the fraction of sources with excess RV noise is $\sim$30\%), when compared to the fraction of sources with excess astrometric noise ($\sim$10\%) \citep{Penoyre_2022a}. This tradeoff between excess radial velocity and astrometric noise among low-mass dwarfs may be driven by several factors. As \cite{Penoyre_2022a} found, lower-mass binaries with orbital periods detectable from \Gaia\ astrometry (that is, $<10$ years) would necessarily reside in smaller orbits. For this reason, the underlying astrometric signal becomes more dominated by noise. However, this same orbital proximity would produce a fractionally \textit{larger} radial velocity signature, potentially explaining some of the discrepancy. On the other hand, low-mass dwarfs are likelier to be active, which contributes significantly to the radial velocity uncertainty \citep{Saar97}. 

\begin{figure}[ht]
\centering
\includegraphics[width=0.49\textwidth]{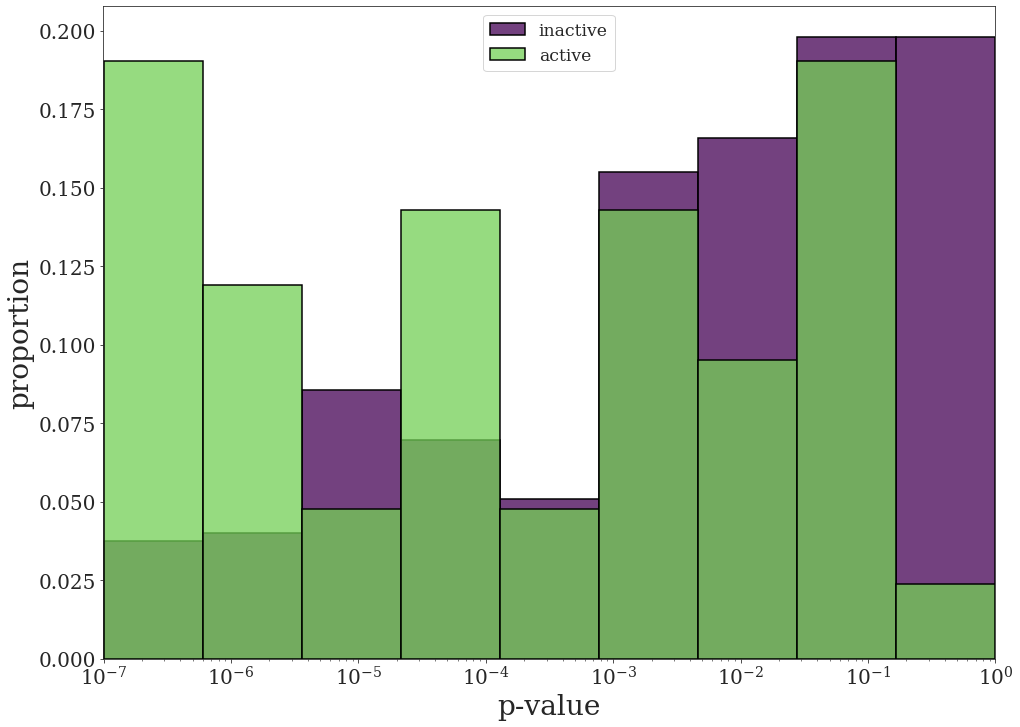}
\includegraphics[width=0.49\textwidth]{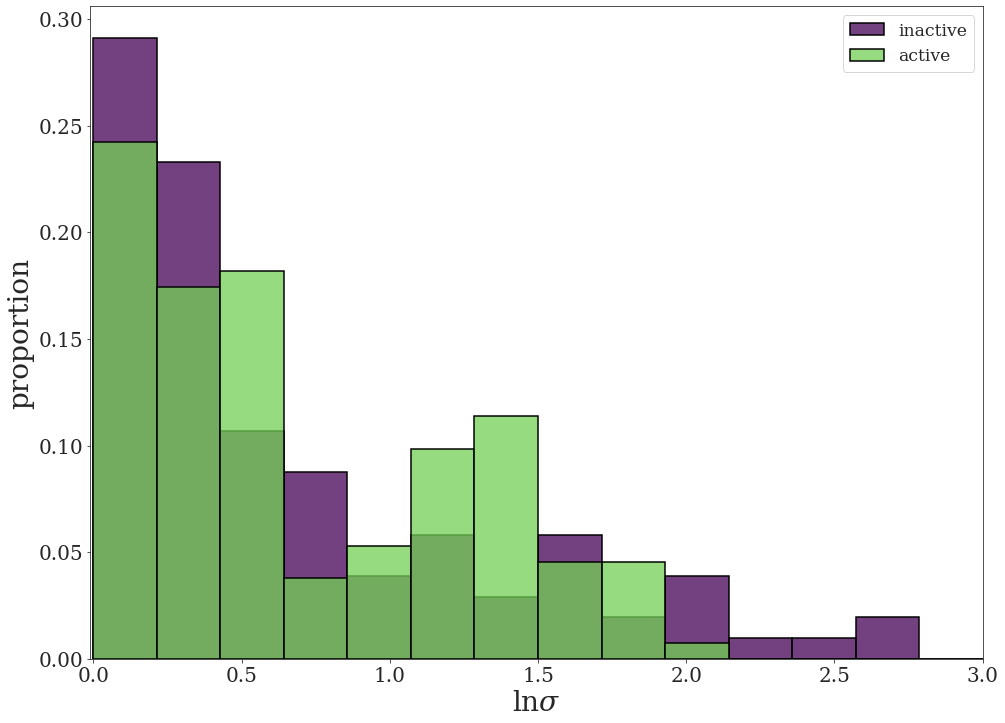}
\vspace{0mm}
\caption{\textit{Top}: A histogram of the $p$-values of 724 M-dwarfs from \cite{newton_2017}, split by H$\alpha$ activity. Active M-dwarfs are more likely to have \textit{excess} RV noise than their inactive counterparts.
\textit{Bottom}: A histogram of the natural log of the per-transit RV measurement uncertainty (see Appendix \ref{sec:appendix-noise}). This value is larger for active M-dwarfs, leading to  more outliers when compared to a random sample of similar stars that includes both active and inactive M-dwarfs.}
\label{fig:thyme}
\end{figure}

We perform a cursory examination of the effect of stellar activity on \paired's $p$-values for cool dwarfs, shown in Figure \ref{fig:thyme}. We consider the $p$-values calculated by \paired\ for a sample of nearby M dwarfs observed by \cite{newton_2017}. In that study, the authors employed H$\alpha$ in emission to classify stars as ``active" versus ``inactive." We do indeed find that active cool stars exhibit higher per-transit measurement uncertainty, when compared to inactive ones. They are therefore more likely to be flagged as ``binaries'' when compared to a sample containing both types. It may be possible to address this more carefully with a sample that includes rotation information (though, per \citealt{newton_2017}, period co-varies with H$\alpha$), and indeed \cite{Penoyre_2022b} already noted the relationship between photometric variability and astrometric noise in the \Gaia\ sample.

By default, \paired\ estimates the semiamplitude $K$ (see Section \ref{sec:calc_radvel}) for stars with $p$-value $<0.01$, a less stringent cutoff than the criterion for ``excessive'' noise (that is, $p<0.001$). The number of stars that satisfy $p$-value$<0.01$ is 5,929,048, or 19\% of the sample. 

\subsection{Comparison with analogous DR3 quantities}
\label{sec:DR3_comparison}

In this Section, we focus upon \paired's findings, when compared against analogous quantities from the \Gaia\ DR3 catalog of sources observed by the Radial Velocity Spectrograph. In the DR3 data release, the fields \rvchisqpval\ and \rvrenormalisedgof\ can be used as potential flags for binarity for about 6 million bright sources \citep{katz_2022,blomme_2022}. As we describe in more detail in Section \ref{sec:intro}, these are two among many ``noise" keywords in the \Gaia\ catalog that have been employed as proxies for binarity in the literature. We compare \paired's $p$-value against \rvchisqpval\ in Section \ref{sec:rvchisqpval}, and against \rvrenormalisedgof\ in Section \ref{sec:rvrenormalizedgof}. 

\subsubsection{Comparison with \Gaia's \rvchisqpval}
\label{sec:rvchisqpval}

\begin{figure}[ht]
\centering
\includegraphics[width=0.49\textwidth]{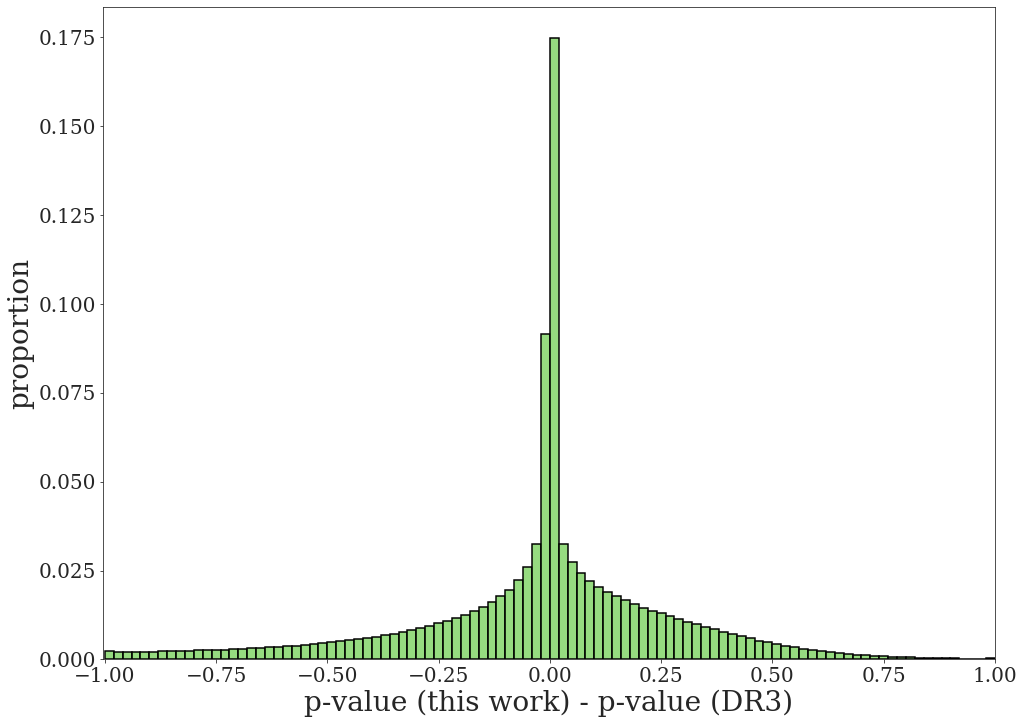}
\caption{Histogram of  difference between $p$-values calculated in this work vs. $p$-values from Gaia DR3. \citep{katz_2022}.}
\label{fig:pval_histogram}
\end{figure}

Subtracting the \paired\ $p$-value from the \rvchisqpval\ field for the same sources, we see general agreement (see Figure \ref{fig:pval_histogram}): for 90\% of sources, there is agreement within 0.1. For cases where our $p$-values do not agree, we attempt to provide context for the disagreement (though there exists an inherent limitation, by virtue of having access only to the RV keywords, and not the epochal measurements). To understand these deviations better, we map the difference in between \paired\ $p$-value and \rvchisqpval\ to the HR diagram in Figure \ref{fig:pval_heatmap}. 

It appears that both Gaia and \paired\ are flagging stars in the correct areas of the color-magnitude diagram, but the lower median \paired\ $p$-values implies more confidence in that RV variability.

The bottom panel of Figure \ref{fig:pval_heatmap} shows this deviation, this time as a function of the number of stars at that color and magnitude. We see the most agreement where there are the most stars: bins with more than 1000 sources tend to converge toward the same median $p$-value in both catalogs. Because \paired\ asks only: ``does this source show excess RV noise compared to typical noise for stars at that color and magnitude?," the number of stars is critical toward establishing ``typical." In more sparsely populated places, any \textit{a priori} assumption placed on radial velocity variability will have outsized influence. Indeed, these more sparsely-populated bins are where \paired\ deviates from \rvchisqpval. 

As mentioned in Section \ref{sec:false-positives}, \paired\ flags stars slightly below the lower Main Sequence much more frequently. We speculate that this difference comes from a combination of effects. The noise model used by Gaia may change for stars that are known to be active, i.e. young stars, rapidly rotating K and M dwarfs. If this is true, assumptions about noise would start to drive the RV model in a way that diverges from the average behavior of similar stars. These sources could also be Main Sequence-white dwarf binaries considering their position slightly blue-ward of the Main Sequence.
\paired\ also flags evolved stars ($B_{P}-R_{P}\sim$1.7, $M_{G}\sim$4) more frequently, which aligns with expectations for the anomalously cool and rotationally enhanced giants that occupy this position  \citep{patton_2024}.

We therefore advise caution, and careful consideration of noise priors, for portions of the HR diagram in which \paired's $p$-values diverge substantially from \Gaia's \rvchisqpval. This includes, among other stellar types, dwarfs stars cooler than about 4000K.

\begin{figure}[h]
\centering
\includegraphics[width=0.49\textwidth]{figures/difference_hr_panel_1.png}
\caption{\textit{Top panel:} Color-magnitude diagram of the difference between $p$-values calculated in this work vs. $p$-values from \Gaia\ DR3 \citep{katz_2022}. Sources for which there is more disagreement between the \paired\ $p$-value and \rvchisqpval\ cluster in locations on the HR diagram that are more sparsely sampled. \textit{Bottom panel:} The difference between \paired\ $p$-value and  \rvchisqpval\ as a function of the number of stars in that portion of color-magnitude space. There is broad agreement  where $N\gtrsim10^{3}$, but where \paired\ is calibrated on fewer stars, our quantification of ``excess'' noise deviates more from DR3.}
\label{fig:pval_heatmap}
\end{figure}

\subsubsection{Comparison with \Gaia\'s \rvrenormalisedgof}
\label{sec:rvrenormalizedgof}

There are multiple keywords for \Gaia\ sources that have historically been employed to identify ``excess" noise in the literature. We summarize in Section \ref{sec:intro} how deviation from a single-star model, whether in astrometry or radial velocity observations, has been applied to cull probable binaries from stellar samples. These diagnostics include the astrometric goodness-of-fit (\verb|GOF_AL|, per \citealt{Evans18}) and the renormalized unit weight error (\textit{RUWE}, per \citealt{Belokurov20}). We consider here the relationship between the \paired\ $p$-value, and the keyword \rvrenormalisedgof. Both terms quantify ``excess" radial velocity noise in slightly different ways. \rvrenormalisedgof is a renormalized unit weight error parameter from the RV time series, which is
larger than unity for variable sources \citep{katz_2022}. Presumably these quantities ought to trend together, with binary sources presenting both \textit{low} $p$-value and \textit{high} \rvrenormalisedgof. The latter is typically applied in the literature via a cutoff, such as e.g. identifying binaries with \rvrenormalisedgof$>$4 \citep{Cao22}. 

\begin{figure}[ht]
\centering
\includegraphics[width=0.49\textwidth]{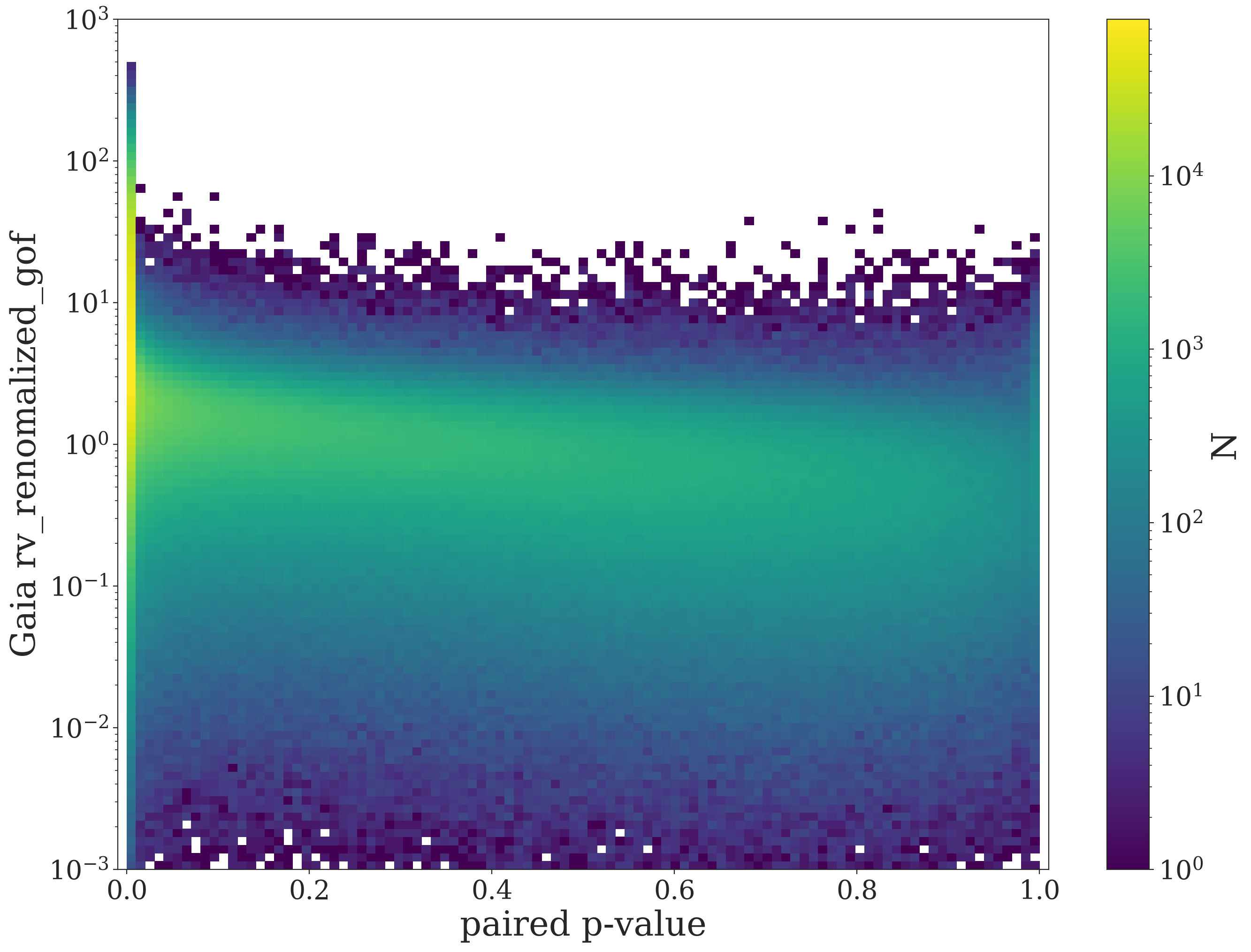}
\includegraphics[width=0.49\textwidth]{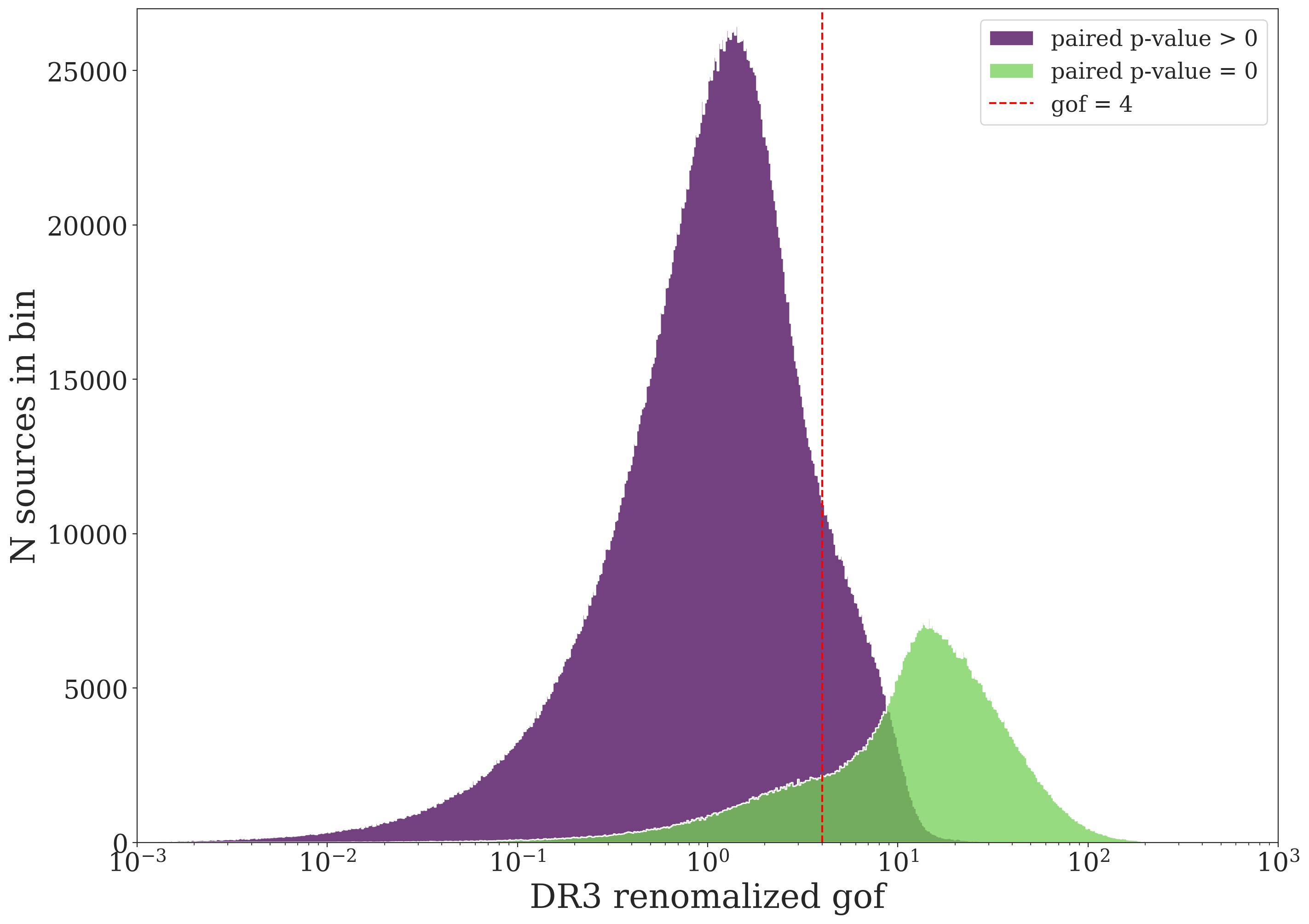}
\vspace{0mm}
\caption{\emph{Top}: A sensitivity map of the \paired\ $p$-value vs. the DR3 field \rvrenormalisedgof. The colorbar corresponds to the number of sources in each bin. \rvrenormalisedgof\ is largely uncorrelated with the \paired\ $p$-value outside the 0 $p$-value bin in this figure.   \emph{Bottom}: The distribution of \rvrenormalisedgof\ for sources with \paired\ $p$-values greater than 0 (in purple) and equal to 0 (in green). sources that our catalog identifies as binaries have a \rvrenormalisedgof\ distribution that peaks well past the \rvrenormalisedgof\ $= 4$ threshold identifies in \cite{katz_2022} as an indicator of RV variability}
\label{fig:gof}
\end{figure}

In Figure \ref{fig:gof} we show the relationship between \paired\ $p$-value and \rvrenormalisedgof, for the $\sim$6 million \Gaia\ RVS sources with published \rvrenormalisedgof\ values. We focus on two regimes: the $0.001<p<1.0$ regime associated with a low likelihood of ``excess" RV noise, and the $p<0.001$ regime associated with high likelihood of ``excess" noise. The top panel shows the trend across the entire $p$-value range. The quantities are only weakly correlated over the range of $p$-values between 0.01 and 1, with high scatter. A source with a \paired\ $p$-value in this range exhibits a typical amount of radial velocity noise when compared to similar stars; such a star likely also furnishes a good fit to a single-star radial velocity model as quantified by \rvrenormalisedgof.  Where the two quantities ought to strongly trend together is actually for the high ``excess" noise regime at $p<0.001$. The bottom panel of Figure \ref{fig:gof} shows the distribution for \rvrenormalisedgof, comparing the two regimes. For higher $p$-values (no evidence of ``excess" noise) we see that \rvrenormalisedgof\ behaves similarly to a normal distribution, peaking at 1. In contrast, the \rvrenormalisedgof\ distribution is offset to a higher value for sources with $p<0.001$, with a mode of 15. We have overplotted the \rvrenormalisedgof$>$4 cutoff. There is not a clear 1:1 correlation for these \rvrenormalisedgof$>$4 with their $p$-value. They trend together on average, but the two quantities diverge in whether ``excess" noise is present approximately 20\% of the time. We see that for $\sim$20\% of the sources where \paired\ identifies excess noise ($p<0.001$), the   \rvrenormalisedgof$<$4, and similarly for $\sim$20\% of sources without excess noise per \paired\, the \rvrenormalisedgof\ $>$4. 
 
\subsubsection{Comparison with \Gaia's identification of non-single-stars and single-lined spectroscopic binary stars}
\label{sec:rvchisqpval-id}

\begin{figure*}[ht]
\centering
\includegraphics[width=0.49\textwidth]{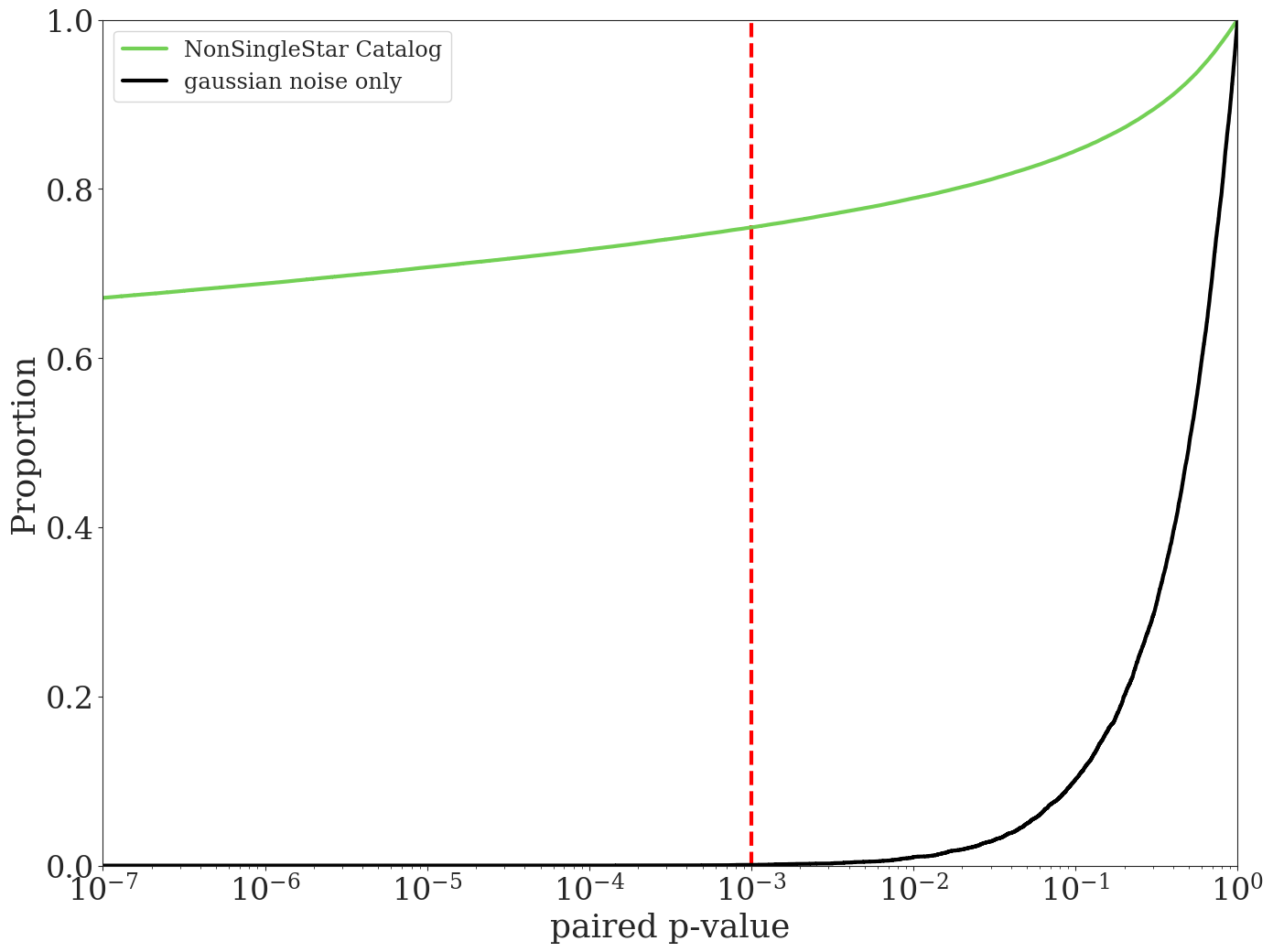}
\includegraphics[width=0.49\textwidth]{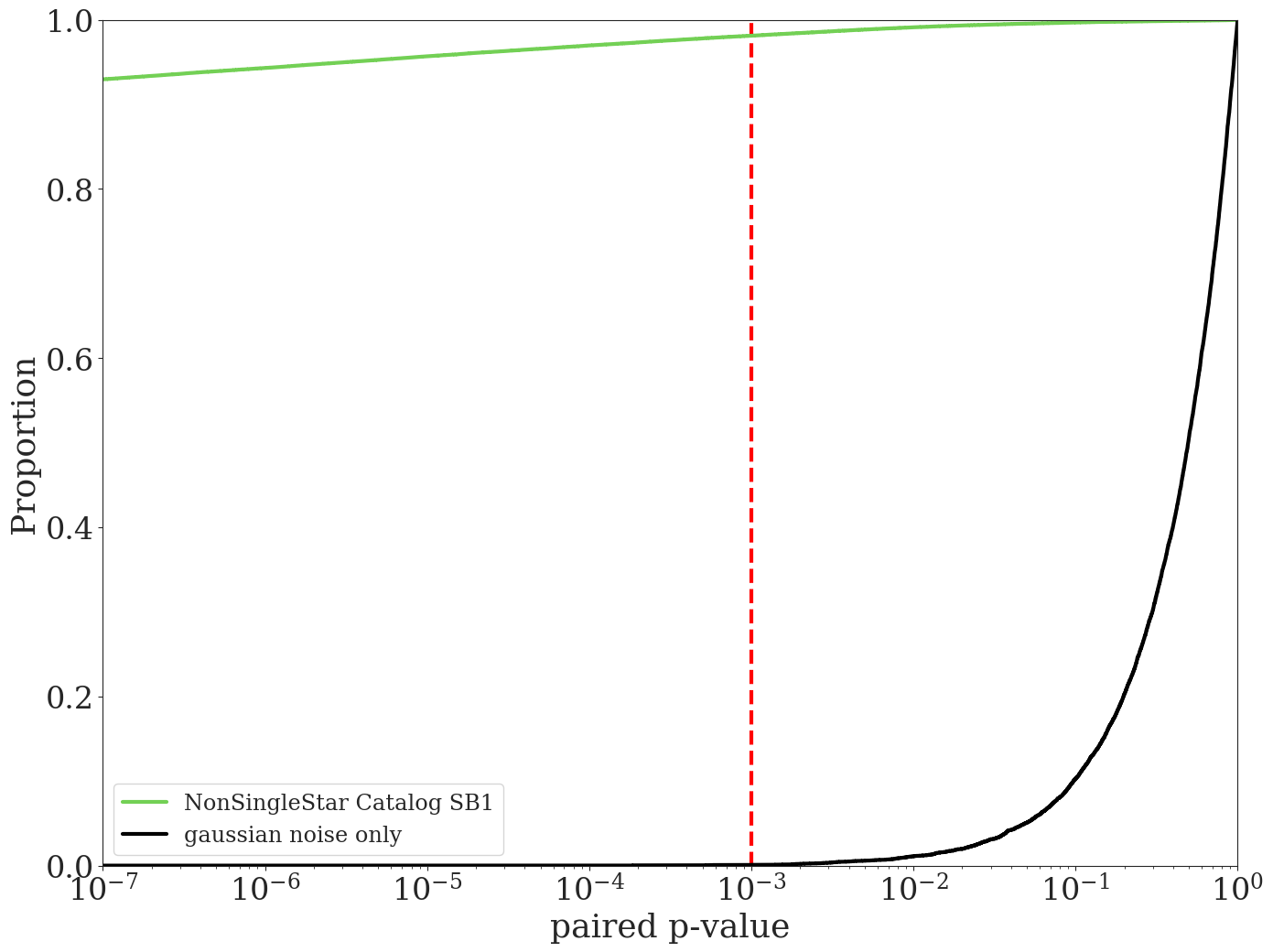}
\vspace{0mm}
\caption{Left: The cumulative density function (CDF) of the  $p$-values of DR3 sources with two-body orbit solutions \citep{gaia_release_2023}. The red dashed line is our $p<0.001$ threshold. The intersection of this line with the CDFs is the proportion of the sample that are flagged as binaries. Right: Identical plot for the only sources flagged as SB1 in the table. \label{fig:TBO_cumulative}}
\end{figure*}

An appropriate first-pass test of this pipeline is a comparison to known binaries in DR3. The \Gaia\ DR3 Catalogue contains $\sim$800,000 sources with orbital solutions of some kind (both orbital elements and trend parameters) for astrometric, spectroscopic, and eclipsing binaries. A significant fraction of the solutions for these binaries derive from astrometry, though some sources also observed by RVS are identified as single-lined spectroscopic binaries while others are simultaneously astrometric and spectroscopic binaries. In the latter case, there are sometimes combined asterometric and radial velocity solutions \citep{gaia_mult}. These solutions reside in the \verb|nss_two_body_orbit| table, and a query for sources with at least 3 RVS observations results in non-single-star orbital models for over 300,000 sources. Though we know that these sources all have at least \textit{some} radial velocity observations, the solutions may rely mostly or almost entirely on astrometry. For this reason, we anticipate that these 300,000 binaries reflect a mixture of orbital inclinations, ranging from edge-on to face-on. \paired\ identifies $\sim$75\% of the sources in the \verb|nss_two_body_orbit| table as binaries, with $p<0.001$.  Figure \ref{fig:TBO_cumulative} shows the distribution in \paired\ $p$-values for the sample as a whole. We infer that some sizeable fraction of the 25\% that are missed by \paired\ is due to a more face-on geometry. In comparison, \paired\ correctly identifies 98\% of the 179,000 sources flagged as single-lined spectroscopic binaries (SB1s) in the \verb|nss_two_body_orbit| table. We conclude that \paired\ is often effective at identifying, from excess RV noise alone, unresolved binaries.  

\subsubsection{Comparison with \Gaia's \semiamplitudeprimary}
\label{sec:DR3_compare_K}
We investigate in Section \ref{sec:rvchisqpval-id} whether \paired\ can correctly identify, by excess RV noise, the $\sim300,000$ binaries with Keplerian orbit solutions \citep[][Gosset et al.\ in prep]{halbwachs_2022}. Beyond their identification, we also test the extent to which \paired\ can recover consistent estimates for the semi-amplitude of the primary. Though we do not have access to the the epochal RVs (which will allow a much better constrained solution), \paired\ ought to show general agreement with their method. In particular, if our inferred semi-amplitudes agree with the \verb|nss_two_body_orbit| sample, it lends credibility to the estimates for similar targets. 

\begin{figure*}[ht]
\centering
\includegraphics[width=0.49\textwidth]{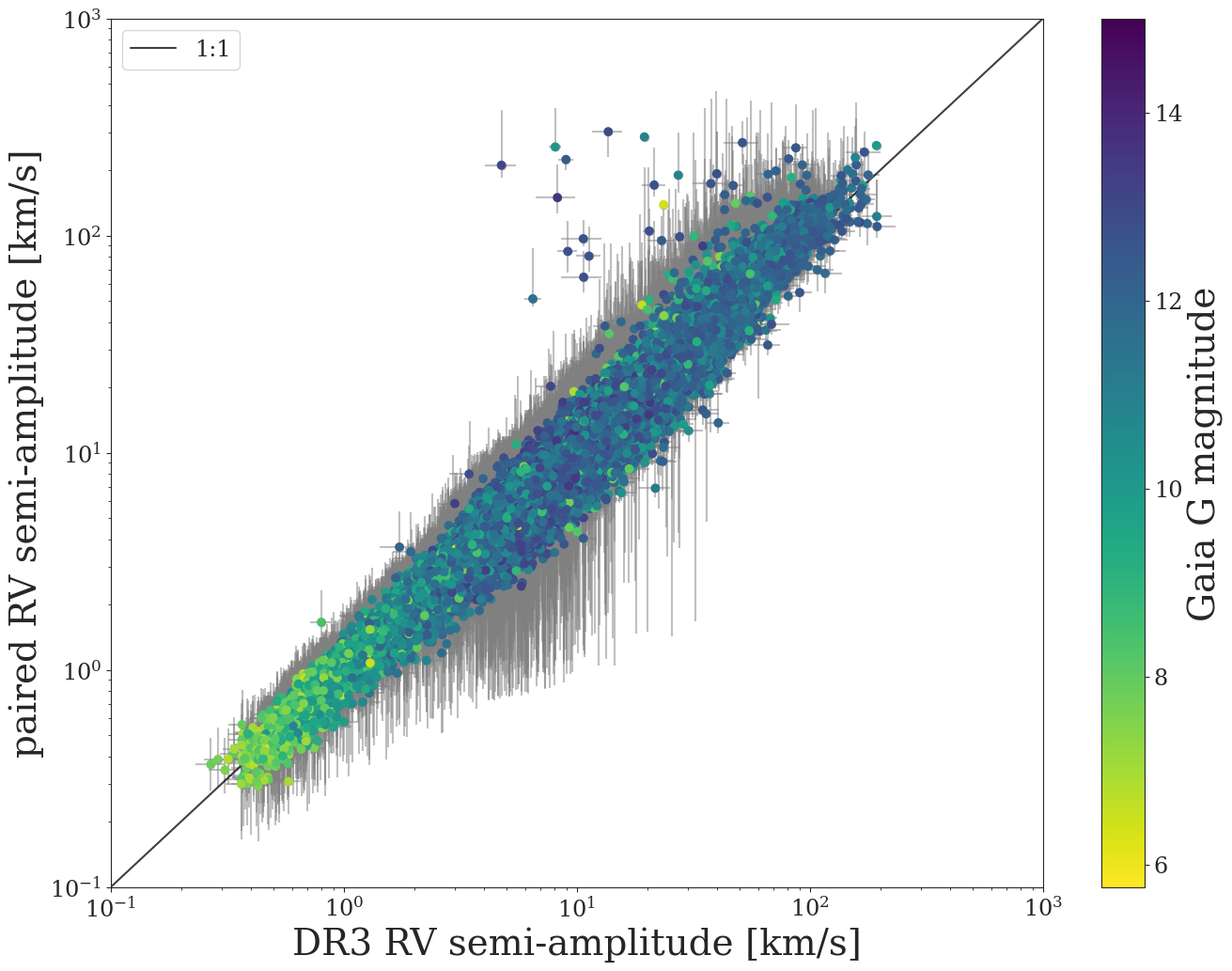}
\includegraphics[width=0.49\textwidth]{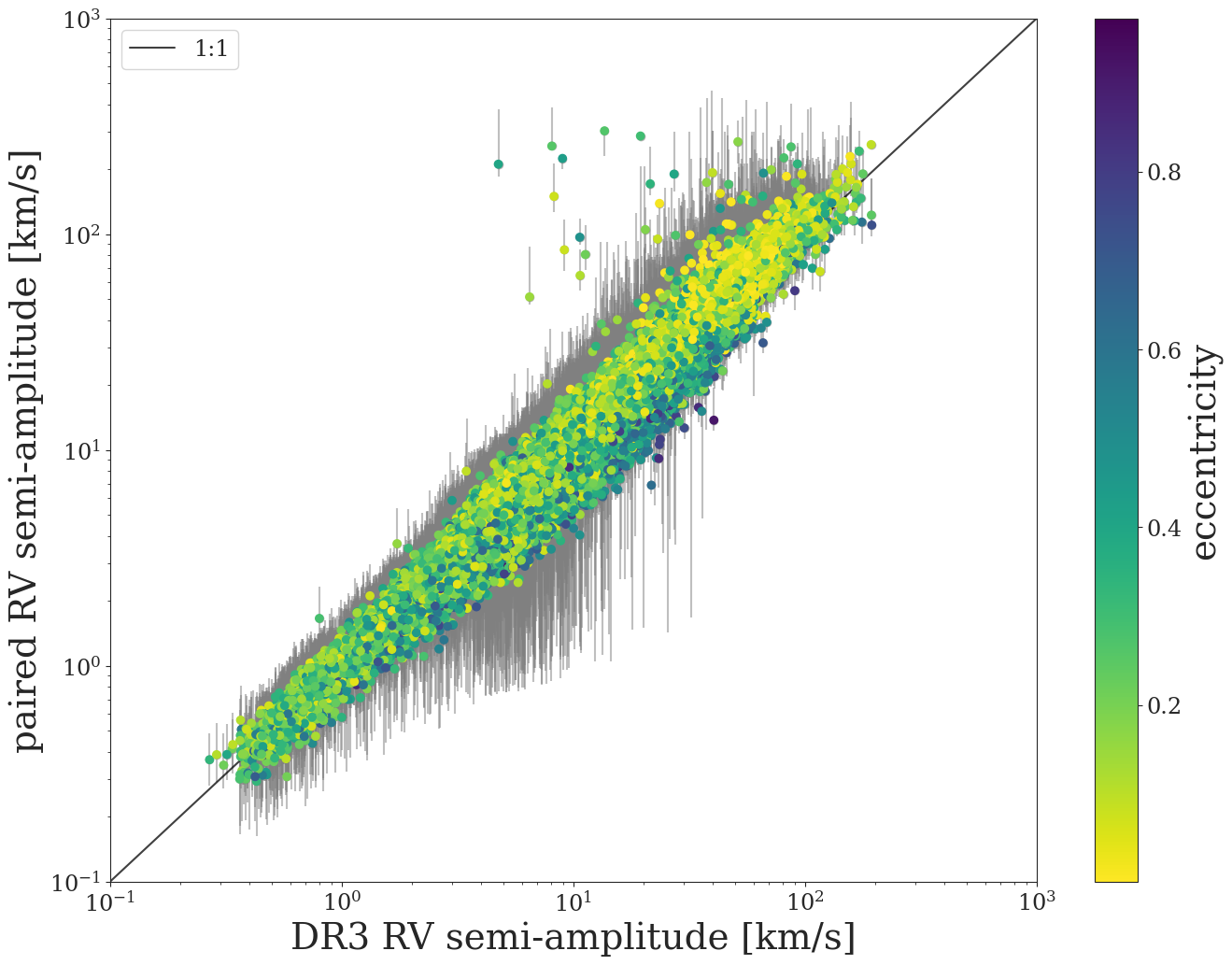}
\vspace{0mm}
\caption{Here we compare our results with 177,410 spectroscopic binary RV semi-amplitudes reported in \Gaia\ DR3. The left panel is the comparison of our estimated semi-amplitude to the reported semi-amplitude colored by magnitude. There is tight agreement across the magnitude range. The right panel is the comparison of estimated semi-amplitudes for the same sources, this time colored by the estimated RV uncertainty for each source. } \label{fig:DR3_K_comparison}
\end{figure*}

In Figure \ref{fig:DR3_K_comparison}, we have made this comparison. In these figures, the plotted sources are from the \verb|nss_two_body_orbit| table sample that have been identified as single-lined spectroscopic binaries with orbital solutions. They all have \paired\ $p$-values lower than our threshold of 0.001, indicating the presence of excess RV noise when compared to stars of similar magnitude and color. The points lie clustered near the 1:1 line, indicating that we recover the correct semi-amplitudes within the 16th and 84th percentile range for most sources. 


\subsection{Validation Against Other Catalogs of Known Binaries}
\label{sec:Binary_comparison}

\subsubsection{When does \paired\ identify known eclipsing binaries?}
\label{sec:paired_EB_samples}

We consider here \paired's ability to identify eclipsing binaries, employing the EB catalogs of the \textit{Kepler} and \textit{TESS} missions. While we hope the application of \paired\ to the \textit{Kepler} and \textit{TESS} stellar samples will be illuminating for planet demographic studies (we comment on this further in Section \ref{sec:conclusion}), we consider these questions outside the scope of this work. For the purpose of validating \paired, we consider only the subset of stars identified as EBs from these transit surveys. 

Known eclipsing binaries (EBs) in the \textit{Kepler} and \textit{TESS} catalogs present an ideal sample for testing \paired's performance. In theory, the transiting geometry of the EB favors its detectability by radial velocity. They comprise the types of edge-on and short-period systems to which \paired's completeness shows high sensitivity (per Section \ref{sec:injection}), and can be confirmed as binaries independent of spectroscopic measurements. While \textit{TESS} stars are brighter on average than \textit{Kepler} stars, eclipsing binaries reside in the part of $\{a,q\}$ space where we retain high recovery rates even for dim ($m_{g}>15$) stars, as shown in Figure \ref{fig:completeness-observed}. Our recovery rate will depend on a combination of factors: the distribution in $\{a,q\}$ space of the detected eclipsing binaries (itself a function of the underlying transit survey sensitivity), our detection efficiency over that range, and the magnitude distribution of the stars.

In collecting the sample of EBs, we cross-match the Kepler EB catalog (containing 2922 sources, \citealt{villanova_eb}) and the TESS EB catalog (containing 4584 sources, \citealt{tess_eb}) with the $\sim$30 million stars with \Gaia\ RVS observations. We winnow the sample slightly by requiring at least 3 radial velocity measurements (that is, ``transits'') from \Gaia. Figure \ref{fig:RV_transits} shows the mean is $\sim$20 transits, and indeed only a tiny fraction are removed by the $N\ge3$ criterion. Of the crossmatched \textit{Kepler} EB sample with at least three transits (comprising 331 EBs)  $73.8\%$ exceed the $p<0.001$ threshold for ``detection'' with \paired. \paired\ performs similarly for the TESS EB catalog \citep{tess_eb}. A \Gaia\ crossmatch in which we again require $N\ge3$ results 458 EBs. Of these, we recover $73.1\%$ as likely binaries. This is in comparison with the approximately 12\% of the Kepler Target List (KTL) and 26\% of Community Target List (CTL) stars, from the full catalog cross-match, that exhibit excessive radial velocity noise at the $p<0.001$ level.

\begin{figure*}[ht]
\centering
\includegraphics[width=0.49\textwidth]{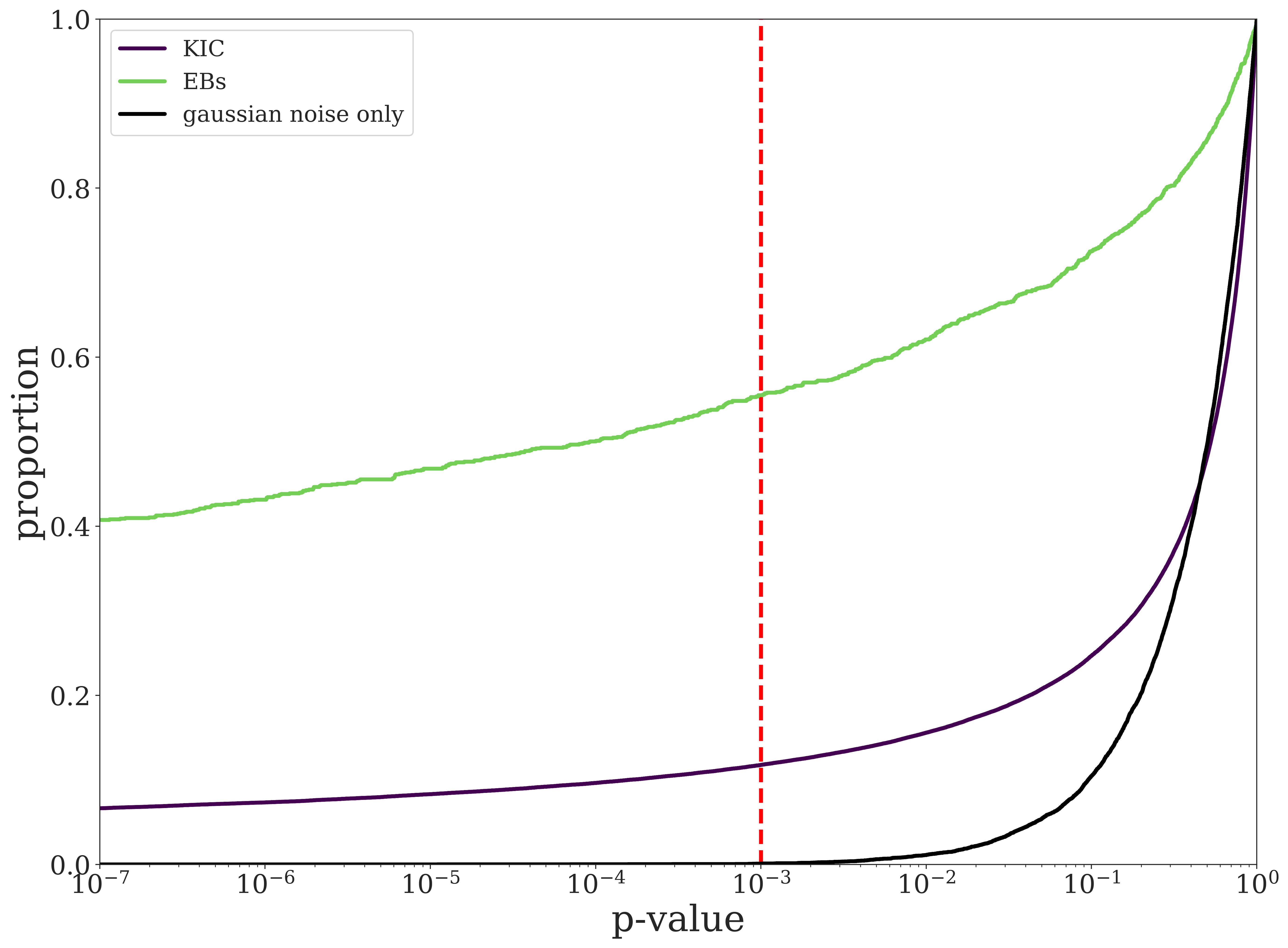}
\includegraphics[width=0.49\textwidth]{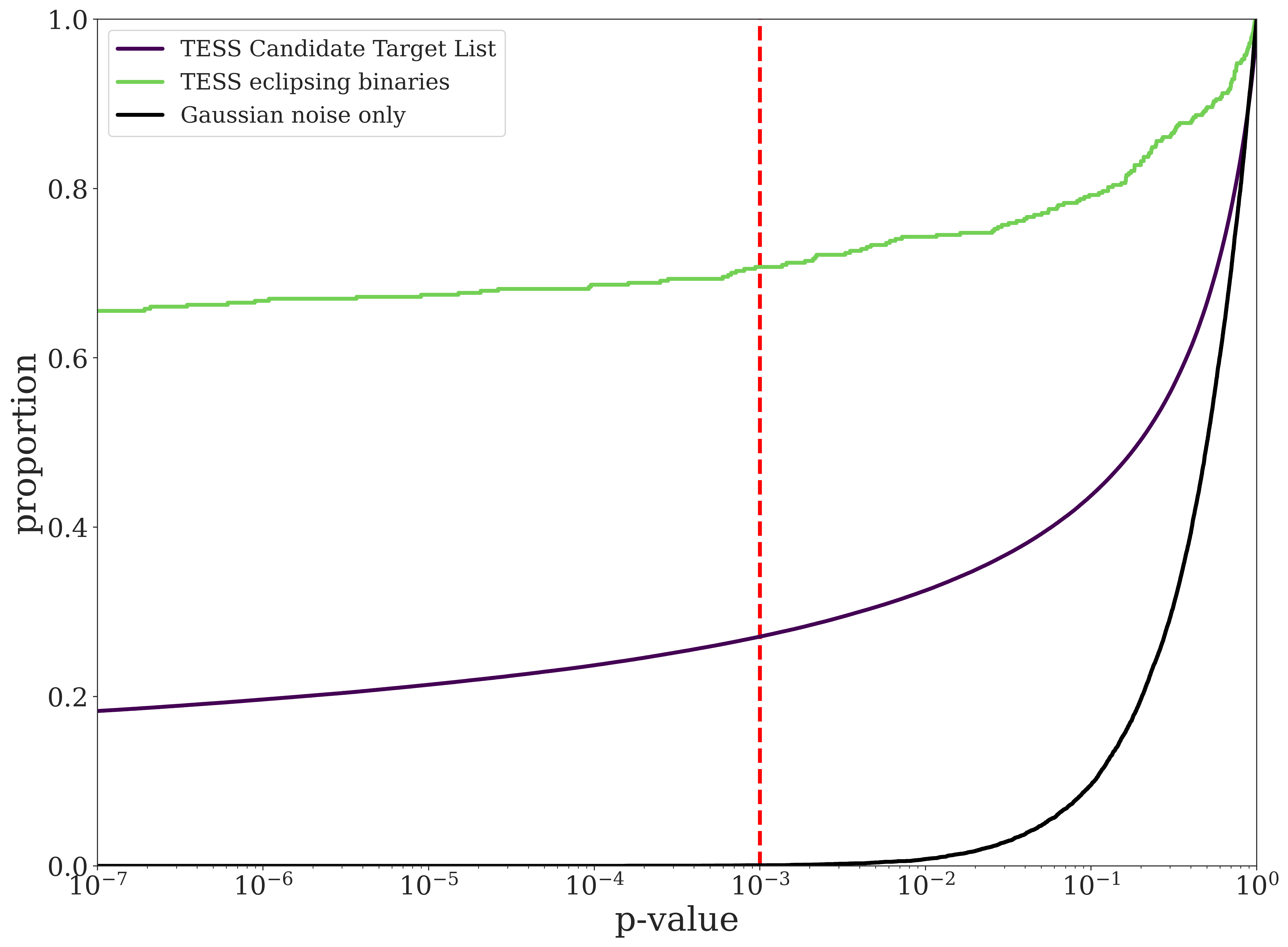}
\vspace{0mm}
\caption{Left: CDF of $p$-values of KTL crossmatch and Villanova Kepler Eclipsing binary catalog \citep{villanova_eb}. The red dashed line is our $p<0.001$ threshold. The intersection of this line with the CDFs is the proportion of the sample that are flagged as binaries. Right: Identical plot for the TESS Community Target List and Villanova TESS Eclipsing binary catalog \citep{tess_eb}. Eclipsing binaries from both catalogs have excess RV noise that meets our threshold in  $\sim 70\%$ of stars, compared to the "field" rate of $\sim 15-20\%$.\label{fig:EB_cumulative}}
\end{figure*}

Figure \ref{fig:EB_cumulative} shows the cumulative density function (CDF) of $p$-values for all of our cross-matched sources from \textit{Kepler} and \textit{TESS}. For a given  $p$-value on the x-axis, the intersection of that vertical line with the CDF indicates the proportion of values that are less than or equal to the selected $p$-value. The $p$-value also reflects the probability that, given our data, a source would exhibit the same level of noise in the idealized Gaussian noise case. 



\subsubsection{When does \paired\ identify known spectroscopic binaries?}
\label{sec:paired_SB_samples}

We also consider \paired's ability to positively identify known spectroscopic binaries.  We perform a cross-match with the SB9 catalog \citep{Pourbaix_2004}. We expect a higher recovery fraction with \paired\ for spectroscopic binaries \textit{a priori} (as opposed to eclipsing binaries), as both samples will be shaped by completeness in a similar way: that is, from radial velocity survey sensitivity, rather than transit survey sensitivity. Indeed, of the 1669 single-lined spectroscopic binaries with at least 3 \Gaia\ RV observations included in \cite{Pourbaix_2004}, we recover $82.8\%$. Intriguingly, this figure is similarly to the $\sim$75\% recovery of binarity in the SB9 sample by \cite{Belokurov20}, using \Gaia's astrometric noise (RUWE) rather than radial velocity noise. One interpretation is preference of SB9 binaries, as with other binaries detected by radial velocity, to reside closer to edge-on geometries. Such edge-on configurations would present relatively smaller signals in astrometry than pole-on configurations, providing one plausible explanation for \paired's slightly higher recovery rate. 


\subsubsection{Comparison with \paired\ radial velocity semi-amplitude against spectroscopic binary catalogs}
\label{sec:compare_RV}

For a sample of spectroscopic binaries in the SB9 \citep{Pourbaix_2004} and APOGEE \citep{Kounkel21} samples, we consider not only whether the binary is detected via \paired's $p$-value, but also whether we recover a consistent radial velocity semi-amplitude. As we describe Appendix \ref{sec:appendix-semiamp}, we calculate the probability distribution for the semi-amplitude from the \Gaia\ radial velocity noise alone. Inferred semi-amplitude posteriors from sparsely sampled data generally have long tails, accommodating the possibility of larger eccentricity values \citep{PriceWhelan17}. This is in comparison to the semi-amplitudes measured from radial velocity surveys with epochal radial velocity observations. 

\begin{figure*}[ht]
\centering
\includegraphics[width=0.7\textwidth]{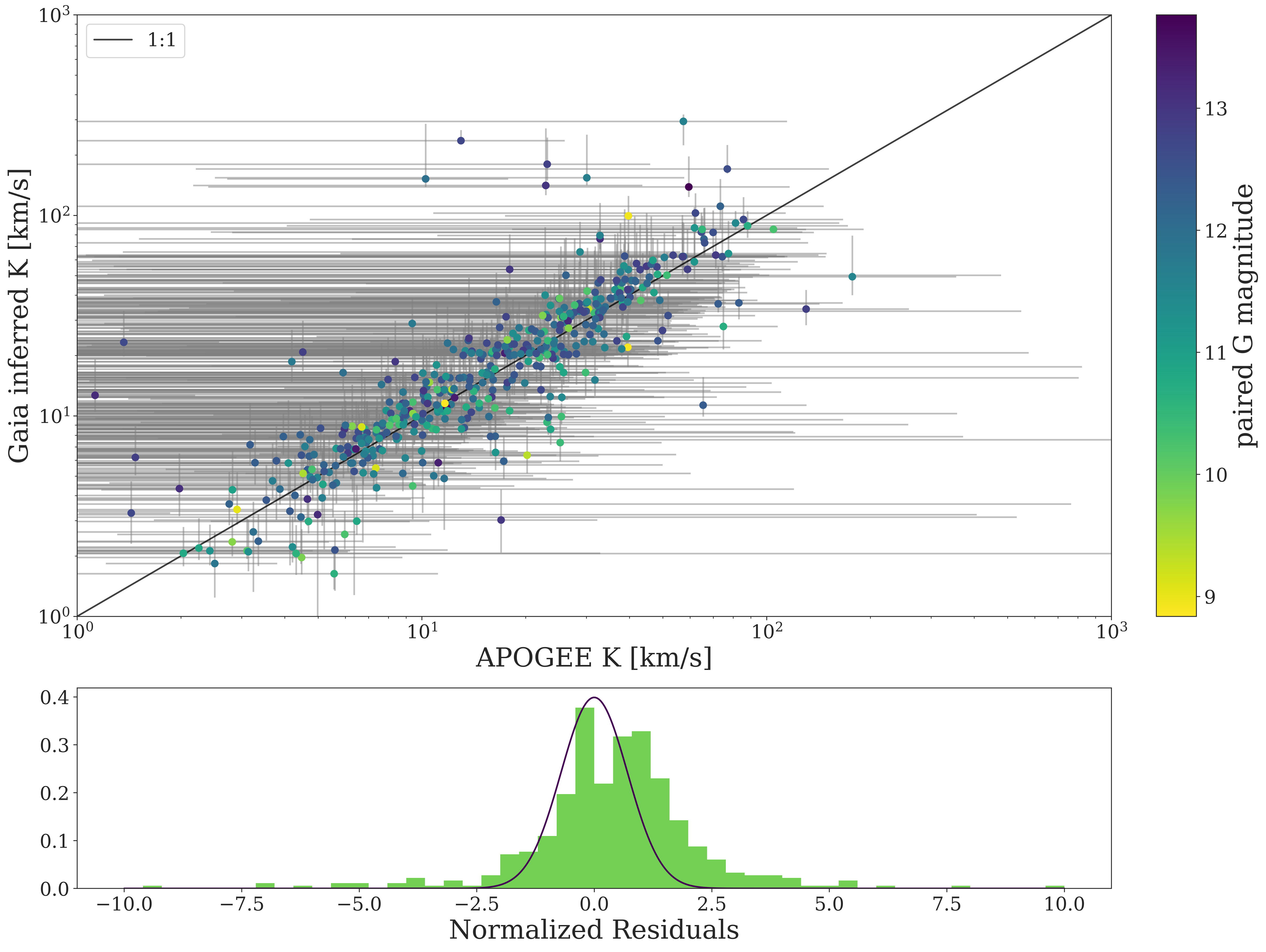}
\includegraphics[width=0.7\textwidth]{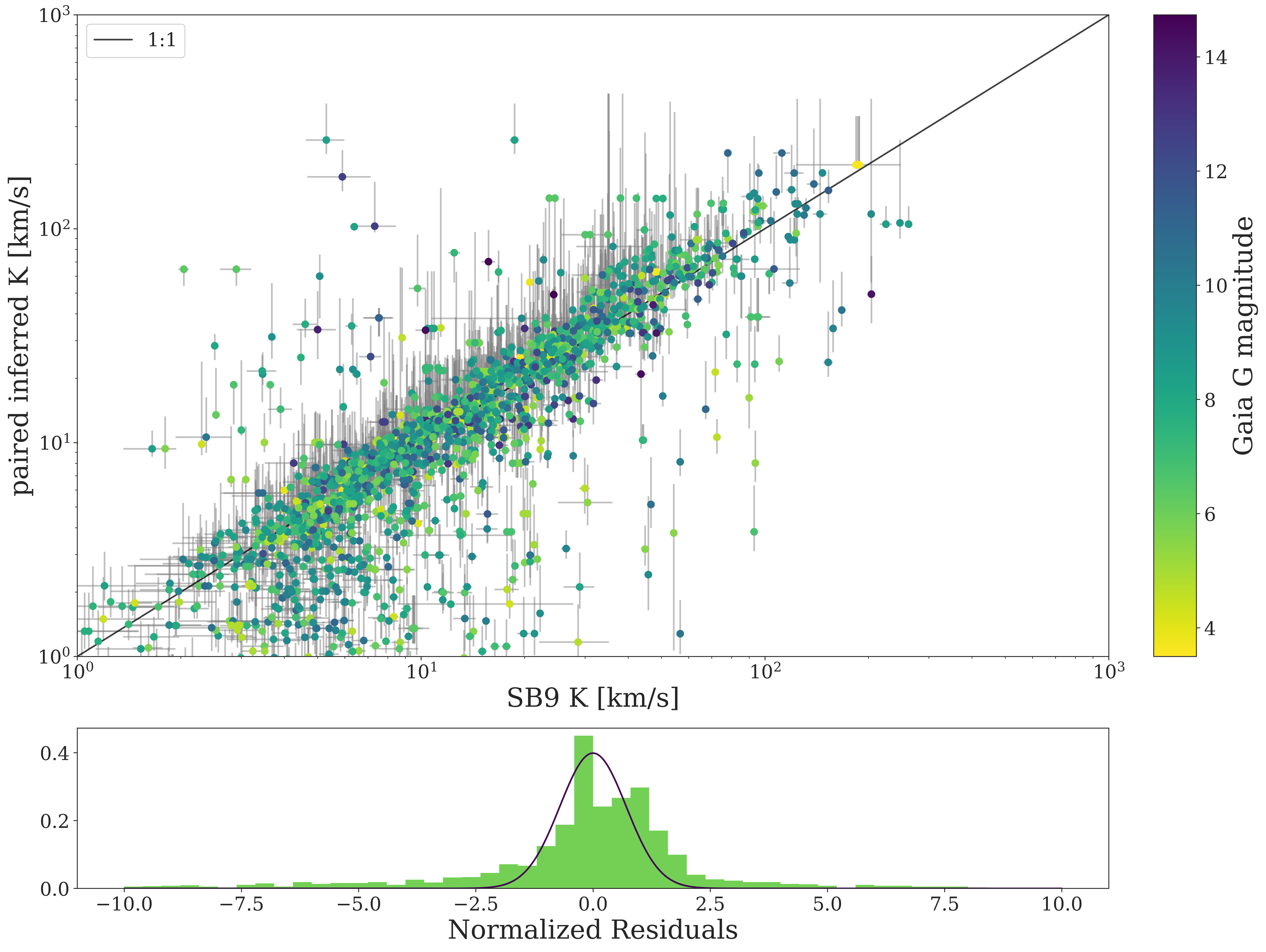}
 
\caption{A comparison between the semi-amplitude $K$ inferred by \paired, with the semi-amplitude inferred from ground-based radial velocity surveys of binaries. \textit{Top panel:} A comparison between the semi-amplitude inferred by \paired, and the value measured with mid-resolution RV survey APOGEE \citep{Kounkel21}. \textit{Bottom panel:} A comparison between the semi-amplitude inferred by \paired, and the value from the spectroscopic binary catalog SB9 \citep{Pourbaix_2004}. There is a tight agreement for high semi-amplitudes that star to loosen as K approaches the \Gaia\ noise floor of $\sim 1$ km/s.}
\label{fig:RV-comparison}
\end{figure*}

The vast majority of the crossmatched RVs are in agreement to 1-$\sigma$ (see Figure \ref{fig:RV-comparison}, \citealt{Price_Whelan_2020}). Our ability to infer semi-amplitude degrades near the \emph{Gaia} RV noise floor of about 1 km/s. Below this floor, the RV signal is indistinguishable from noise. We note a slight offset visible in the residuals from the 1:1 line. This offset may potentially be attributable to a known effect with magnitude, where the inferred radial velocity from \Gaia\ is  overestimated for dimmer sources (0 for $G_{\textrm{RVS}}\sim$9, up to 500 m/s for $G_{\textrm{RVS}}\sim$11.75) \citep{Katz19, Tsantaki22}. \cite{katz_2022} posited that the offset may be attributable to charge trapping in the CCD, and while DR3 was produced with updated calibration to address the offset, an additional iteration of correction ($\sim$500 m/s $G_{\textrm{RVS}}\sim$14) proved necessary, to bring \Gaia\ median velocities into consistency with APOGEE stars. We are aware, therefore, of existing systematics in the radial velocity measurements at the $\sim$500 m/s level, which may underly \paired's observed overestimation of $K$ at the $\sim$500 m/s level. We emphasize that this offset is still smaller than the typical 1 km/s precision for even the brightest sources.

\section{Conclusion}
\label{sec:conclusion}

In this manuscript, we describe a statistical framework called \paired\ for identifying stellar multiplicity using radial velocity noise from \Gaia\ DR3. The resulting $p$-values are extracted from two of the three presently available measurements for a subset of \Gaia\ target stars: the error associated with the mean radial velocity and the number of spectroscopic observations (or ``transits''). From these values, we construct a statistically robust mechanism for determining consistency between the reported radial velocity error for a source, and the distribution of radial velocity error for stars of similar color and magnitude. Where inconsistent, we generally assume that the excess noise is indicative of additional orbital reflex motion of the source (due to a companion). We demonstrate that this assumption is borne out in general when compared to other benchmarks of binarity: samples of known eclipsing binaries, the clustering of ``likely" binaries on the binary sequence of the HR diagram, and existing radial velocity catalogs. However, we also consider ``false positives," where a source shows excess RV noise when compared to other stars of similar magnitude and color, but not due to \textit{bona fide} Keplerian motion. The nature of these false positives probably varies across the HR diagram in a similar manner to sources of ``excess'' astrometric noise from \Gaia\ \citep{Belokurov20, Penoyre_2022a}. A comparison between \paired's $p$-value and the analogous \rvchisqpval\ product from DR3 highlights the extent to which assumptions about the underlying stellar noise model matter. When \paired\ is supplied with $\gtrsim10^{3}$ stars at a given magnitude and color, the ``excessive noise" $p$-value nearly always replicates \rvchisqpval\ (with the exception of some stars on the sub-Main Sequence). It is only where we have less empirical knowledge about the stellar population ($<10^{3}$ stars) that prior assumptions about stellar noise matter most, and where we show deviation from DR3.  

Despite our advertised caution, this catalog is intended to be useful wherever some idea of the close binary rate of stars is useful. In the field of exoplanet surveys, specific possible use cases include:
    
    \begin{itemize}
        \item Vetting of planet candidates and identification of false positives
        \item Comparison between distributions of planet- and non-planet-hosting stars
        \item Identification of circumbinary planet candidates
        \item Direct detection of sub-stellar-mass companions in the brown dwarf and giant planet mass range, and  
        \item Examination of confirmed planetary systems with anomalous radial velocities. 
    \end{itemize}

We have confined ourselves in this manuscript to a description of the \paired\ machinery, and leave the application of it to exoplanetary studies to future work. However, we note that a sizeable fraction of both the \textit{Kepler} and \textit{TESS} samples have been observed by RVS, so that \paired\ $p$-values are immediately available for them. The crossmatch with \textit{Kepler} contains 96,748 matches out of nearly 200,000 Kepler target stars, down to a \Gaia\ G magnitude of 15.83. The crossmatch with \TESS\ contains 854,700 out of approximately 9.5 million stars in the CTL, down to a \Gaia\ G magnitude of 15.81. This comprises roughly 50\% of the surveyed stars for \textit{Kepler}, and 10\% of the surveyed stars for \textit{TESS}. 

Beyond applications to exoplanets, \paired\ can also be helpful for studying rates of Main Sequence/compact object binaries, or Main Sequence/brown dwarf binaries. For example, \cite{Andrews19} estimated the extent to which \Gaia\ might be sensitive to these binaries, though an exercise to establish ``completeness" \textit{per se} would require an injection-and-recovery exercise that characterizes ``detection>" We aim for \paired\ to provide such a tool. This list of suggested use cases is not complete, and we are hopeful that the catalog will be broadly useful to the community interested in stellar populations.

\begin{acknowledgments}
This work was performed in part through the Pre-Doctoral program at the Flatiron Institute's Center for Computational Astrophysics.

Q.C.'s work was supported by the National Aeronautics and Space Administration under Grant No. 80NSSC21K1841 issued through the Future Investigators in NASA Earth and Space Science and Technology program.

This work has made use of data from the European Space Agency (ESA) mission \Gaia\ (https://www.cosmos.esa.int/ gaia), processed by the Gaia Data Processing and Analysis Consortium (DPAC; https://www.cosmos.esa.int/web/gaia/ dpac/consortium). Funding for the DPAC has been provided by national institutions, in particular the institutions participating in the Gaia Multilateral Agreement.

This research has made use of the Exoplanet Followup Observation Program website (ExoFOP), which is operated by the California Institute of Technology, under contract with the National Aeronautics and Space Administration under the Exoplanet Exploration Program. 

This research has made use of the NASA Exoplanet Archive, which is operated by the California Institute of Technology, under contract with the National Aeronautics and Space Administration under the Exoplanet Exploration Program.

This work made use of the gaia-kepler.fun crossmatch database created by Megan Bedell.

We are thankful for helpful discussions with Natalia Guerrero, Chris Lam, Veselin Kostov, Kaitlin Kratter, Elisa Quintana, and Jamie Tayar. 
\end{acknowledgments}

\paragraph{Software}
    astropy \citep{2013A&A...558A..33A, 2018AJ....156..123A}, numpy \citep{harris2020array}, matplotlib \citep{Hunter:2007}, pandas \citep{reback2020pandas, mckinney-proc-scipy-2010}

\appendix

\section{$\chi^2$ $p$-values as a detection metric}
\label{sec:appendix-chi2}

For each RV target, the \Gaia\ catalog reports the median RV, an estimate of the error on this median, and the number of individual RV measurements (called ``transits'') that were used in this estimate.
In our analysis, we ignore the median RV and instead only consider the RV error and number of transits.
For targets brighter than $G_\mathrm{RVS} = 12$, the RV error $\epsilon_k$ for a target $k$ is computed as \citep{Katz19, katz_2022}
\begin{eqnarray}\label{eq:rv-error-def}
	{\epsilon_k}^2 = \left(\sqrt{\frac{\pi}{2\,N_k}}\,s_k\right)^2 + 0.113^2
\end{eqnarray}
where
\begin{eqnarray}
	\label{eq:sample-var}
	{s_k}^2 = \frac{1}{N_k-1}\sum_{n=1}^{N_k} \left(v_{k,n} - \bar{v}_k\right)^2
\end{eqnarray}
is the sample variance of the individual RV measurements $v_{k,n}$.
In Equation \ref{eq:sample-var}, $n=1,\cdots,N_k$ indexes the $T_n$ transits for target $n$ and
\begin{eqnarray}
	\bar{v}_k = \frac{1}{N_k}\sum_{n=1}^{N_k} v_{k,n}
\end{eqnarray}
is the mean RV.
In all that follows, it is much simpler to reason about the sample variance ${s_k}^2$ instead of the error on the median ${\epsilon}_n$, but that isn't a problem because we can compute
\begin{eqnarray}
    \label{eq:sample-var-comp}
	{s_k}^2 = \frac{2\,N_k}{\pi}\left({\epsilon_k}^2 - 0.113^2\right)
\end{eqnarray}
for any target, using only values available in the catalog: $\epsilon_k$ and $N_k$.
Now, for a given target $k$, in the absence of any excess radial velocity signal, and if we know its per-transit RV measurement uncertainty $\sigma_k$, we would expect the quantity
\begin{eqnarray}
	{\xi_k}^2 &=& \frac{(N_k - 1)\,s_k^2}{{\sigma_k}^2}
	\label{eq:chi-sq-samp}
\end{eqnarray}
to be chi-squared distributed with $N_k - 1$ degrees of freedom.
Therefore, we can compute the $p$-value for the measured RV error as
\begin{eqnarray}
	\mathrm{Pr}(\xi^2 > {\xi_k}^2\,|\,\mathrm{null}) &=& \int_{{\xi_k}^2}^\infty \chi^2 (\xi^2;\,N_k-1) \dd \xi^2 \quad,
    \label{eq:chi2-$p$-value}
\end{eqnarray}
where the ``null hypothesis'' is that the target is a single star.
This $p$-value quantifies, under our measurement uncertainty model, how frequently we would expect to measure the catalog value of the RV error if that target were a single star.
Therefore, targets with strong evidence for anomalously large RV errors will have low $p$-values, and we can use this value for identifying potential binary star systems or other types of statistical outliers.
The \Gaia\ DR3 catalog also includes a similar $p$-value measurement for targets brighter than $G_\mathrm{RVS} = 12$, but we use our own measurements of this quantity here for several reasons.
First, since we have access to the full pipeline we can characterize the expected detection efficiency for our method, using injection and recovery tests, as discussed in Section \ref{sec:injection}.
This product is important for potential future applications of our method to detailed population studies.
Furthermore, since we have developed our own pipeline for estimating the per-transit RV measurement uncertainty for each target, we can use that to infer the expected semi-amplitude of any candidate binary systems as discussed in Section \ref{sec:compare_RV} and Appendix \ref{sec:appendix-semiamp}.

Although the formal mathematical motivation for this model doesn't directly generalize to the RV reduction methods used for targets fainter than $G_\mathrm{RVS} = 12$, we apply exactly the same method to the fainter targets in the catalog.
When we compare our results to literature binary catalogs in Section \ref{sec:Binary_comparison}, we find consistent results without systematic effects at fainter magnitudes.
This suggests that this methodology can be usefully applied across a wide range of magnitudes, significantly increasing the accessible parameter space of our results.

\section{Estimating the per-transit RV measurement uncertainty}
\label{sec:appendix-noise}

\begin{figure}[htbp]
    \centering
    \includegraphics[width=0.7\textwidth]{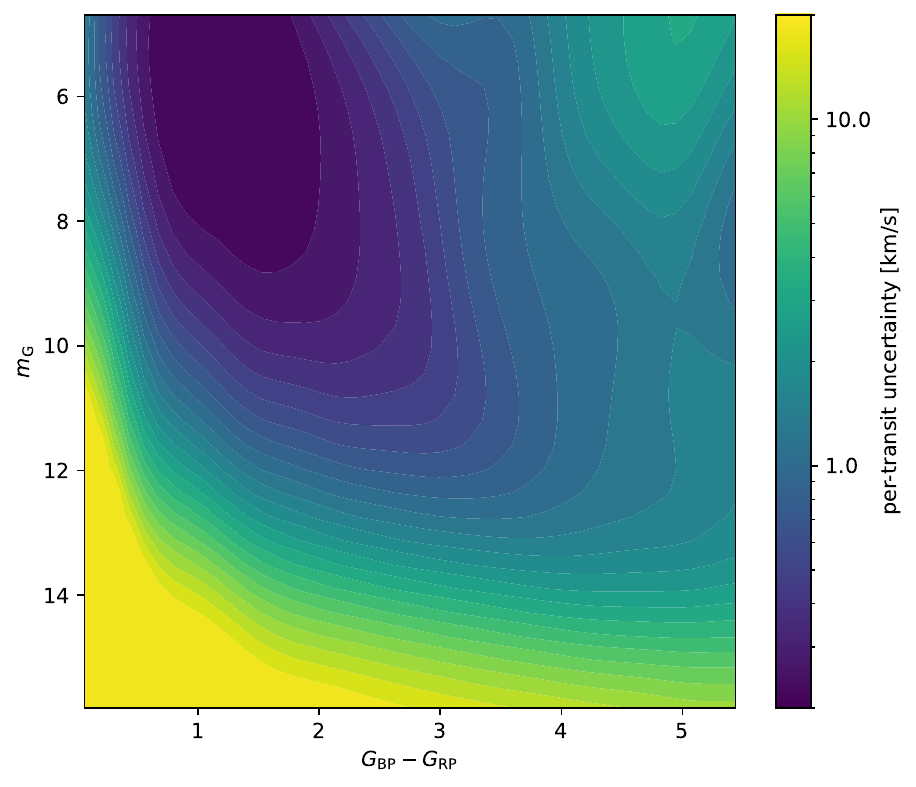}
    \caption{The expected per-transit RV measurement uncertainty as a function of \Gaia\ color and magnitude, computed based on the \Gaia\ DR3 RV error measurements and an iterative sigma clipping method.
    This figure shows these estimates smoothed and interpolated using a maximum likelihood Gaussian Process model in two dimensions. 
    \label{fig:noise_model}}
\end{figure}

In the previous discussion, we assumed that the per-observation RV measurement uncertainty $\sigma_n$ was known.
This is not, however, a number that we have direct access to in the official \Gaia\ data products.
Instead, we calibrate this empirically from the population-level RV errors in the catalog, as a function of the target's apparent magnitude and observed color.
For a fine two-dimensional grid in magnitude and color\footnote{We use rectangular bins in color and magnitude with a bin width of 0.1~mag and 0.4~mag respectively.}, we estimate the per-transit RV uncertainty using an iterative sigma-clipping method, described below.
This procedure assumes that the RV measurement uncertainty is the same for all targets with similar colors and apparent magnitudes and that none of these bins include mostly non-single stars.
For each bin in our grid, we compute the RV sample variance $s^2$ using Equation \ref{eq:sample-var-comp} for all the targets in the bin.
Then, we make an initial estimate of the squared measurement uncertainty by taking the sample median of the $s^2$ values.
In the limit of a large sample of targets known to be single stars, we would expect this to be a well-calibrated estimator, but in the real sample, some targets will be outliers, namely the binaries.
Therefore, we compute the $\chi^2$ $p$-value for each target using Equation \ref{eq:chi2-$p$-value} and the current estimate of the measurement uncertainty and remove targets with $p$-values less than $0.001$.
Then using this updated sample, we iterate this clipping procedure until the sample converges.
Using simulated catalogs, we find that this procedure can reliably recover the true observational RV uncertainty to better than $5\%$, even if as many as $90\%$ of the targets in a bin are binaries.
To decrease the bin-to-bin variance across the grid, we smooth and interpolate the estimated uncertainty values using a maximum likelihood Gaussian Process model in two dimensions.
We then use this Gaussian Process model to compute the expected uncertainty for all the RV targets in the \Gaia\ DR3 catalog.
Figure \ref{fig:noise_model} shows the final estimate of the per-transit RV measurement uncertainty across the full range of \Gaia\ color and apparent magnitude.

\section{Estimating RV semi-amplitudes}
\label{sec:appendix-semiamp}

We can even go one step further than just detecting candidate binary systems, and also estimate the RV semi-amplitude of these systems based solely on the reported RV error and the number of transits.
To make this estimate, we derive the expected sampling distribution for the sample variance ${s_k}^2$ (Equation \ref{eq:sample-var}) under a model for the excess RV signal.
In our case, we want to infer the orbital properties of any binary systems, so we use a Keplerian orbit as our underlying RV model \citep{Lovis:2010}
\begin{eqnarray}
	\mu_k(t;\,\theta_k) &=& v_{0,k} + K_k\,\left[\cos(f_k(t) + \omega_k) + e_k\,\cos \omega_k\right] \quad,
	\label{eq:rv-model}
\end{eqnarray}
but the following derivation could be used more generally.
In Equation \ref{eq:rv-model}, the parameters $\theta_k$ are the usual Keplerian orbital elements for system $k$---the mean RV $v_{0,k}$, the RV semi-amplitude $K_k$, the orbital eccentricity $e_k$, and the argument of periastron $\omega_k$---and $f_k(t)$ is the true anomaly computed at time $t$.

We can evaluate Equation \ref{eq:rv-model}, at observation times $\{t_{k,n}\}_{n=1}^{N_k}$, but in our case, we don't actually \emph{know} the observation times\footnote{The Gaia Observation Forecast Tool (\url{https://gaia.esac.esa.int/gost/}) predicts the transit times for astrometric measurements, and we expect the RV measurements will be a subset of these transits, but for most sources that information is not enough to substantially improve our inferences.}, so in practice, we will need to marginalize those out, but for now, let's proceed as if we do know them.
For a given set of parameters $\theta_k$, it can be shown that the sampling distribution for our scalar ${\xi_k}^2$ from above (Equation \ref{eq:chi-sq-samp}) is a non-central $\chi^2$ distribution with $N_k$ degrees of freedom, and non-centrality parameter
\begin{eqnarray}
	\lambda &=& \sum_{n=1}^{N_k} \left[\frac{\mu_k(t_n;\,\theta_k) - \bar{\mu}_k}{\sigma_k}\right]^2
	\label{eq:ncx2-nc-param}
\end{eqnarray}
where
\begin{eqnarray}
	\bar{\mu}_k &=& \frac{1}{N_k}\sum_{n=1}^{N_k} \mu_k(t_n;\,\theta_k)
\end{eqnarray}
is the sample mean of the model evaluated at the observation times.
This sampling distribution defines a likelihood function for the parameters $\theta_k$, and it is the key ingredient of our method.

Having defined this likelihood function, we model every \Gaia\ RV target as a bound binary system, and infer a probabilistic constraint on the orbital parameters under this model.
To make these inferences, we use prior importance sampling: we draw samples of all the orbital parameters and nuisance parameters from the priors listed in Table \ref{tab:priors}\footnote{Note that the mean RV offset, $v_{0,k}$ in Equation \ref{eq:rv-model}, is not sampled because all factors of $v_{0,k}$ cancel in Equation \ref{eq:ncx2-nc-param}.
Then, using these samples, we evaluate the non-central chi-squared likelihood function using Equation \ref{eq:ncx2-nc-param}, and re-weight samples with this likelihood.
This produces an estimate of the posterior probability density for each of the physical parameters listed in Table \ref{tab:priors}, but in practice, this procedure does not significantly constrain any of the parameters except RV semi-amplitude.
Since the likelihood is not very informative in most dimensions, and the fit is low dimensional, this prior sampling approach is more computationally efficient than a method like Markov chain Monte Carlo.
We parallelize the likelihood calculation using the JAX numerical computing library \citep{jax2018github}, so the computational cost for each target is much less than a second.}
In Section \ref{sec:compare_RV} and Section \ref{sec:false-positives}, we discuss the performance of these inferences when compared to catalogs of spectroscopic binaries from the literature and simulations respectively.

\begin{deluxetable}{ccc}
\centering
\tablecolumns{3}
\tablecaption{The prior densities used when inferring the RV semi-amplitude for each \Gaia\ RV target. \label{tab:priors}}
\tablehead{\colhead{Parameter} & \colhead{Prior} & \colhead{Comment}}


\startdata
$\ln K / \mathrm{km\,s^{-1}}$ & $\mathcal{U}(\ln 0.05,\,\ln 500)$ & RV semi-amplitude \\
$\ln P / \mathrm{day}$ & $\mathcal{U}(\ln 1,\,\ln 800)$ & orbital period \\
$\phi$ & $\mathcal{U}(-\pi,\,\pi)$ & orbital phase \\
$e$ & $\mathcal{U}(0,\,0.9)$ & orbital eccentricity \\
$\omega$ & $\mathcal{U}(-\pi,\,\pi)$ & argument of periastron \\
$\{ t_n / \mathrm{day} \}$ & $\mathcal{U}(0,\,\Delta t_k)$\tablenotemark{a} & observation times \\
\enddata

\tablenotetext{a}{$\Delta t_k$ is the observational baseline for the target $k$ as published in the \Gaia\ DR3 column \texttt{rv\_time\_duration}.}

\end{deluxetable}

\newpage
\bibliography{gaia_paper.bib}
\bibliographystyle{aasjournal}

\end{document}